\documentclass[a4paper,twocolumn,accepted=2026-05-26]{quantumarticle}
\pdfoutput=1

\usepackage[utf8]{inputenc}
\usepackage[toc,page]{appendix}
\usepackage{amsfonts}
\usepackage[dvips]{graphicx}
\usepackage{color}
\usepackage{amsmath}
\usepackage{graphicx}
\usepackage{amssymb}
\usepackage{hyperref}
\usepackage{color}
\usepackage{mathrsfs}
\usepackage{isomath}
\usepackage{amsthm}
\usepackage{epstopdf}
\usepackage{txfonts}
\usepackage{dsfont}
\usepackage{ulem}
\usepackage{physics} 
\usepackage{tikz} 

\PassOptionsToPackage{sort&compress}{natbib}
\usepackage[numbers]{natbib}

\newcommand{\be}{\begin{equation}}
\newcommand{\ee}{\end{equation}}
\renewcommand{\ketbra}[1]{\ensuremath{| #1 \rangle \langle #1 |}}
\newcommand{\kb}[2]{\ensuremath{| #1 \rangle \langle #2 |}}
\newcommand{\comment}[1]{}

\newcommand{\mean}[1]{\ensuremath{\langle{#1}\rangle}}

\renewcommand{\emph}[1]{{\it #1}}

\newcommand{\Bth}{{\boldsymbol \theta}}
\newcommand{\COV}{{\mathcal Cov}}
\newcommand{\Cov}{{\Gamma}}
\newcommand{\Covv}[2]{{\Gamma}_{#2}[#1]}
\newcommand{\Qfish}[2]{\ensuremath{\mathcal F^Q_{#2}[#1]}}
\newcommand{\Qfishrho}[1]{\ensuremath{\mathcal F_Q[#1]}}

\newcommand{\Th}{\hat\Bth}
\newcommand{\rhoth}{\varrho_{\Bth}}

\renewcommand{\var}[1]{\ensuremath{(\Delta #1)^2}}
\newcommand{\ex}[1]{\ensuremath{\left\langle{#1}\right\rangle}}

\renewcommand{\vec}[1]{\boldsymbol{#1}}

\newcommand{\id}{\openone}

\newcommand{\probez}{\ket{\Psi_{\vec 0}}}

\newcommand{\thetaMoM}{\hat \theta_{M}}

\newcommand{\ignore}[1]{}

\usepackage{cleveref}
\crefname{equation}{Eq.}{Eqs.}
\crefname{observation}{Obs.}{Obs.}
\crefname{lemma}{Lemma}{Lemmata}
\crefname{proof}{Proof}{Proofs}
\creflabelformat{proof}{#2proof#3}
\crefname{remark}{Remark}{Remarks}
\crefname{prop}{Proposition}{Propositions}

\begin{document}

\title{Characterizing resources for multiparameter estimation of SU(2) and SU(1,1) unitaries}

\author{Shaowei Du}
\affiliation{State key Laboratory of Artificial Microstructure and Mesoscopic Physics, School of Physics, Frontiers Science Center for Nano-optoelectronics, Peking University, Beijing 100871, China}
\author{Shuheng Liu} \email{liushuheng@pku.edu.cn}
\affiliation{State key Laboratory of Artificial Microstructure and Mesoscopic Physics, School of Physics, Frontiers Science Center for Nano-optoelectronics, Peking University, Beijing 100871, China}
\author{Frank E. S. Steinhoff} 
\affiliation{Federal University of Mato Grosso, FCT/DCET, 78060-900, V\'arzea Grande, MT, Brazil}
\affiliation{University of Bras\'ilia, International Center of Physics, 70910-900, Bras\'ilia, DF, Brazil}
\affiliation{University of Bras\'ilia, Institute of Physics, 70910-900, Bras\'ilia, DF, Brazil}
\author{Giuseppe Vitagliano} 
\affiliation{Vienna Center for Quantum Science and Technology, Atominstitut, TU Wien,  1020 Vienna, Austria}

\begin{abstract} 
We analyze the task of estimating a multi-parameter unitary belonging to the $SU(2)$ or $SU(1,1)$ groups, in a two-bosonic-mode scenario and investigate the scaling of the precision in terms of the total particle number. For the $SU(2)$ case, the total particle number is conserved by the evolution and we discuss optimal states in fixed-$n$ subspaces, identifying eigenstates of $J_z^2$ as useful resources, even allowing simultaneous Heisenberg precision scaling for all three parameters. In the $SU(1,1)$ case instead, the conserved quantity is the particle number difference between the two modes, and we identify useful probe states in the sector with an equal number of particles in the two modes. These states are analogous to the $SU(2)$ case and would also allow simultaneous Heisenberg precision scaling for all three parameters. 

We then consider the more pragmatic scenario of an estimation via expectation values of time-evolved observables, which we restrict to be the first two moments of the generators. We analyze the maximal precision achievable in this setting and we find that the twin-Fock state emerges in both the $SU(2)$ and the $SU(1,1)$ cases as the only one potentially allowing Heisenberg scaling for the estimation of two out of the three parameters. As a complement, we also consider other probe states with fluctuating number of particles, with measurements restricted to quadratic expressions in the mode operators. In this scenario,  simultaneous Heisenberg scaling in multiple parameters seems mostly forbidden, with the only exception being an input two-mode squeezed state for the estimation of a two-parameter $SU(2)$. This extends to the multiparameter scenario the well-established intuition that the performance of a $SU(2)$ interferometer can be enhanced by a prior $SU(1,1)$ operation.
\end{abstract}

\maketitle

\section{Introduction}

Since the early days of quantum theory, interferometers have been employed to demonstrate nonclassical features of quantum systems. The Mach-Zehnder Interferometer (MZI) reveals, among other foundational aspects, wave-particle duality, uncertainty relations and complementarity~\cite{wp1,wp2,wp3,wp4}. The interferometers designed by Hanbury Brown and Twiss and by Hong, Ou and Mandel \cite{hbt,hom,aspect}, reveal higher-order optical correlations and phenomena such as particle bunching and anti-bunching, demonstrating that a quantum description of electromagnetic radiation is unavoidable. Furthermore, the observation that inputting quantum states of light can enhance the performance of interferometers has triggered an enormous amount of investigation on what in current terminology can be called quantum-enhanced metrology, which represents nowadays an entire field of research, as well as one of the most promising immediate applications of quantum technology. In fact, current applications of quantum sensing range from atomic clocks~\cite{atomic_clocks}, magnetometry~\cite{magnetometry}, imaging~\cite{ALBARELLI2020126311}, gravitational wave detectors~\cite{ligo2011gravitational,aasi2013enhanced,ligo,virgo}, and are even reaching the realm of high-energy physics~\cite{chou2023quantumsensorshighenergy}. Quantum metrology~\cite{giovannetti2004quantum,giovannetti2006quantum,giovannetti2011advances} can be thus identified as one of the most important modern developments of interferometry~\cite{parisMetrev09,pezze2014quantum,TothApellaniz2014,Demkowicz_Dobrza_ski_2015,DegenReinhardCappellaroRev2017,schnabel2017squeezed,pezzerev18}.  

The estimation of a single parameter with quantum resources has been extensively investigated in recent times, in different paradigms and also in connection with various notions coming from classical statistics, such as the Cram\'er-Rao bound introducing the Fisher information, the Bayesian estimation, and the geometry of the parameter space~\cite{BraunsteinCaves1994}. In \cite{yurke86}, Yurke, McCall and Klauder established a formulation of interferometers in terms of the groups $SU(2)$ and $SU(1,1)$, introducing a group-theoretic approach to metrology. They showed that an MZI could be fully described in terms of $SU(2)$ operations, while squeezing operations could be understood in terms of $SU(1,1)$. Previously, Caves \cite{caves_1981} discovered that squeezed states can enhance the performance of an MZI  beyond the so-called Standard Quantum Limit (SQL). The standard description of spin and two-level systems in terms of $SU(2)$ allowed the derivation of analogous interferometers for these systems and many important concepts in quantum information theory such as spin-coherent states and spin-squeezing were derived in related works~\cite{KitagawaUeda1993,HilleryMlodinow93,Wineland1994Squeezed}. These concepts were also later related to uncertainty relations~\cite{HilleryMlodinow93,BrifMann1996,Brif_1996,brif1996high,Maccone2020squeezingmetrology} and entanglement~\cite{KitagawaUeda1993,Pezze2009Entanglement,Hyllus2012Fisher,TothEntanglementMetrologyFisher12,GessnerPezzeSmerzi16,Fadel_2023}.

In a modern approach to quantum metrology, resources are analyzed in terms of input states, which are characterized via the {\it quantum Fisher information}, providing asymptotic precision limits for phase estimation, and measurements that allow to saturate such limits on the given input state, at least in the limit of a large number of independent repetitions~\cite{parisMetrev09,pezze2014quantum,TothApellaniz2014,Demkowicz_Dobrza_ski_2015,DegenReinhardCappellaroRev2017,schnabel2017squeezed,pezzerev18}. For $SU(2)$ interferometers in which the particle number is fixed and conserved, e.g., in Ramsey schemes with atomic ensembles, typical input states considered are given by spin-squeezed states~\cite{KitagawaUeda1993,HilleryMlodinow93,Wineland1994Squeezed} (or generalizations thereof~\cite{He2011Planar}). Similarly, so-called NOON states~\cite{dowling2008quantum,Leibfried2004Toward}, or twin-Fock states~\cite{Lucke2011Twin,Zou2018Beating,TothApellaniz2014}, have also been considered as useful resources, and investigated from the point of view of their entanglement~\cite{Sorensen2001Entanglement,Hyllus2010Entanglement,Hyllus2012Entanglement,Lucke2014Detecting,TothApellaniz2014,Vitagliano2017Unpolarized,Vitagliano2018Planar,Friis2019,FadelVitagliano_2021}. Many such states have also been the focus of recent experiments~\cite{pezzerev18,Leibfried2004Toward,Wasilewski2010Quantum,Gross2010Nonlinear,Riedel2010Atom-chip-based,Lucke2011Twin,Gross2012Spin,Ockeloen2013Quantum,Muessel2014Scalable,Hammerer2010Quantum,Vitagliano2018Planar}. 

Allowing $SU(1,1)$ operations, which makes the total number of particles not conserved, has also been extensively explored for improving the performance of MZIs, and in that case typical input states considered are, for example, mode-squeezed states~\cite{Gerry2000,pezzesmerziPRL2008,GAIBA2009934,Anisimovetal2010,pezzesmerzi2013,hu2016enhanced,Youetal2019,DuKongetal2022} or cat-like states~\cite{JooMunroSpiller2011,Liu_2016,Chao-Ping_2016,fadel2024quantummetrologycontinuousvariable}. Furthermore, there are also works that consider the estimation of parameters generated by $SU(1,1)$ evolutions, i.e., squeezing coefficients, mainly in the framework of quantum metrology with Gaussian resources~\cite{Li_2014,PinalPRA2013,FriisPRA2015,Sparaciari_15,SparaciariOlivaresParis,SafranekFuentes2016,Rigovaccaetal2017,Andersonetal2017,Gong_2017,Nichols_2018,Bakmou_2020,sorelli2023gaussian}.

Recent years have seen an ever-increasing interest in analyzing the estimation of multiple parameters at once~\cite{TothApellaniz2014,szczykulska2016multi,ALBARELLI2020126311,liu2020quantum}, building upon the abstract observations of Matsumoto~\cite{Matsumoto_2002} concerning the ultimate precision limits of such an estimation based on the Quantum Fisher Information matrix (QFIM). In that case, for the saturation of the QFIM bound there is a crucial difference between the cases in which the generators of the phase-imprinting dynamics commute or not~\cite{Ragy_2016,pezzeetalPRL2017multi,gessnerPRL18,AlbarelliFrielDatta2019,Gorecki2020optimalprobeserror,Demkowicz_Dobrza_ski_2020,Sidhuetal2021}. Several concrete multiparameter scenarios have been considered in recent literature, which include unitary phase imprintings from commuting subsets of $SU(d)$ generators~\cite{Baumgratz_Datta2016}, or non-commuting ones belonging to the $SU(2)$ group~\cite{Vaneph_2013,JingLiuXiongWang2015,Bouchard:17,ChryssomalakosCoronado2017,GoldbergJames2018,Martin2020optimaldetectionof,HouEtAl2020,Z_Goldberg_2021,Gorecki2022,YangRuAnWangZhangLi2022,piotrak2023perfect}, also with group-invariant or parametrization-invariant figures of merit~\cite{AcinetalPRA2001,ChiribellaDarianoPerinottiSacchi,ChiribellaDarianoSacchiPRA05,GoldbergSanchezSotoFerrettiPRL2021}. Furthermore, multiparameter estimation has been considered also in the Gaussian framework~\cite{Nichols_2018,Bakmou_2020,GenonietalPRA2013,Li_2020}.

In general, abstract sensitivity bounds have been explored from the point of view of the QFIM~\cite{yuen1973multiple,Matsumoto_2002,AlbarelliFrielDatta2019,Gorecki2020optimalprobeserror,Demkowicz_Dobrza_ski_2020,liu2020quantum,Sidhuetal2021}, more general bounds~\cite{GessnerSmerzihierarchiesPRL2023} or different, more general frameworks~\cite{Demkowicz_Dobrza_ski_2020}, including the Bayesian approach~\cite{Kay1993FSSP_Estimation,jaynes2003probability,trees2007bayesian}. Moreover, the theoretical saturation of the QCRB~\cite{Ragy_2016,pezzeetalPRL2017multi,YangPangZhouJordan2019} has been investigated, as well as more pragmatic approaches based on signal-to-noise ratios of observables, generalizing the idea of spin-squeezing coming from the error-propagation formula~\cite{gessnerPRL18,Gessner2019Metrological,gessnerNatComm20}.

In this paper, we follow some of these approaches and investigate the case in which the generators of the unitary phase imprinting do not commute, but have some structure, and in particular form a Lie algebra, which is a scenario that has been partly investigated in literature \cite{twamley} also by using group-theoretic tools~\cite{kahn,Imai_2007,GoldbergSanchezSotoFerrettiPRL2021}.

Concretely, we investigate the three-parameter estimation of $SU(2)$ and $SU(1,1)$ unitaries, in the neighborhood of a reference point located at the origin, which are also relevant for practical applications, emphasizing aspects that have potential generalization to the $SU(d)$ and $SU(p,q)$ groups, which we mention as an outlook in the end. We consider the typical scenario of an encoding with a two-bosonic-mode system and, in order to characterize resources for maximal sensitivity, we consider different sectors of pure states that are labelled by group invariants. In the $SU(2)$ case this invariant is just the total particle number, which is the usual resource considered in literature (and related to the energy of the system) and is used to distinguish {\it shot-noise} precision scaling, achieved by usual $SU(2)$-coherent states, versus {\it Heisenberg scaling}. In that case, we then consider the sectors with fixed total particle number and identify as a useful class of states the eigenstates of $J_z^2$, which also contains states that allow reaching Heisenberg scaling for all three phases. However, by then restricting to measurements of the first two moments of the $su(2)$ generators, we observe that the best performance is achieved by the twin-Fock state, which allows Heisenberg scaling only in two of the three phases. 

In the $SU(1,1)$ case, the group invariant is given by the particle number difference between the two modes and $SU(1,1)$-coherent states are generated starting from the vacuum state and essentially correspond to two-mode squeezed vacuum states, which allow Heisenberg precision scaling in the single-parameter case. Analyzing group-invariant sectors, we find that states that have an equal number of particles in the two modes contain, analogously to the $SU(2)$ case, useful states for the three-parameter estimation, including some that allow reaching Heisenberg scaling in all the three parameters. Once again, when we restrict the measurements to be only of the first two powers of the generators, we find that the twin-Fock state remains as the only case allowing the estimation of two parameters simultaneously with Heisenberg scaling. As a comparison we also analyze the performance of further typical states, in particular not having a fixed number of particles, such as Gaussian or cat-like states, and observe that among those only the two-mode squeezed state allows Heisenberg scaling for more than one parameter in a setting with measurement resources limited to quadratic mode operator combinations. From our analysis, we also conclude that in both the $SU(2)$ and $SU(1,1)$ cases the reaching of Heisenberg precision scaling is generally possible only by measuring combinations of mode operators of order higher than fourth.

The work is organized as follows. In \cref{sec:methods} we give an overview of multiparameter estimation strategies based on the quantum Fisher information paradigm, as well as related techniques based on the method of moments. In \cref{sec:HWintermezzo}, we illustrate the main ideas with the Heisenberg-Weyl group of unitaries, generated by two canonically-commuting quadratures. In \cref{sec:SU2multi} we present a detailed account of our results for the multi-parameter estimation of $SU(2)$ unitaries, characterizing the usefulness of $J_z^2$ eigenstates as probe states, that include paradigmatic states like the NOON state and the twin-Fock state. We investigate such states both from the (asymptotic) point of view of the quantum Fisher information matrix, as well as in the more pragmatic scenario of estimation via second-order moments of the $su(2)$ generators. In \cref{sec:SU11multi}, we do a similar analysis for the $SU(1,1)$ case, identifying probe states analogous to the $SU(2)$ case and investigating their performance with estimations using the method of moments with again second-order power of the $su(1,1)$ generators. In \cref{sec:indefN} we discuss states with fluctuating number of particles, including estimation within the Gaussian framework and with cat-like states. In \cref{sec:conclusions} we give our closing statements and perspectives on future developments. In the appendices, we elaborate further on the technicalities of the method of moments and provide more details for the $SU(2)$ case, including aspects related to the different parametrizations.

\section{Methods}\label{sec:methods}

\subsection{Introduction to multiparameter estimation}

We consider the task of estimating a vector of phases $\Bth=(\theta_1,\dots,\theta_K)$ encoded with a unitary evolution on a probe state:
\be
\rhoth = U(\Bth) \varrho_0 U^\dagger (\Bth) ,
\ee
where $\varrho_0$ is the initial state and $U(\Bth)$ is the unitary encoding. We consider a {\it local} estimation problem, in which the phases are considered in an infinitesimal neighborhood of a given point $\vec \theta$. To estimate such a vector of phases one needs to make measurements, which in quantum mechanics are described by a Positive Operator Valued Measure (POVM), i.e., a set of positive operators $\{ M_q \}$ with $M_q\geq 0$ and $\sum_q M_q=\id$, while $q$ labels the outcomes. Here we assume that the vector of parameters can be associated to real numbers that are prior to and independent of the measurements, which is often called the {\it frequentist} approach in literature, and is suitable for our local estimation task. An alternative, more general approach, which is typically termed Bayesian, is usually employed for a global estimation, when a {\it prior} probability distribution $p(\vec \theta)$ is assigned to each point $\vec \theta$ and is then updated to a {\it posterior} $p(\vec \theta| \vec q)$ after some measurements with outcomes $\vec q$ are performed~\cite{Kay1993FSSP_Estimation,jaynes2003probability,trees2007bayesian}.

Thus, collecting statistics with $\nu$ independent measurements, one in principle samples a probability distribution
\be
{\rm p}({\bf q} | \Bth) = \prod_j {\rm p}(q_j | \Bth) = \prod_j \tr[M_{q_j} \rhoth] ,
\ee
where ${\bf q} =(q_1, \dots ,q_\nu)$ is a set of $\nu$ outcomes, and the outcome probability depends on the encoded parameters $\Bth$. Then, one has to find a good vector of estimators $\hat\Bth(\bf q)$, possibly {\it unbiased}, i.e. $\mathbb{E}_{\Bth}[\hat {\Bth}] =\Bth$, the accuracy of which is most commonly quantified via its covariances 
\be
[\COV(\Th)]_{ij}  = \mathbb{E}_\theta[\hat \theta_i \hat \theta_j ] - \mathbb{E}_\theta[\hat \theta_i] \mathbb{E}_\theta[\hat \theta_j] ,
\ee
where we used the notation $\mathbb{E}_\Bth[f(\vec q)]=\sum_{\vec q} f(\vec q) \, {\rm p}(\vec q |\Bth)$ for the expectation value of a function of the outcomes\footnote{In case the estimator is biased, the appropriate figure of merit becomes the {\it mean square error}, defined as $\mathrm{MSE}[\hat\theta]=\mathbb{E}_\theta[(\hat\theta-\theta)^2]$.}. 

\subsection{Quantum Cram\'er-Rao bound and its saturation}

In general, lower bounds can be found as matrix inequalities for the covariance matrix, the most commonly used of which is the celebrated Cram\'er-Rao bound:
\be
\COV(\Th ) \geq \mathcal F^{-1}(\Bth)/\nu ,
\ee
where the matrix on the right hand side is called {\it Fisher information matrix}
\be
\mathcal F_{ij} := \mathbb{E}_\theta[(\partial_{\theta_i} \ln {\rm p}({\bf q} | \vec \theta) )(\partial_{\theta_j} \ln {\rm p}({\bf q} | \Bth))] .
\ee
In turn, the classical Fisher information matrix can be optimized over all POVMs, so to get the additional bound~\cite{Matsumoto_2002,Baumgratz_Datta2016}
\be\label{eq:qCRboundmulti}
\COV(\Th) \geq \mathcal F^{-1}(\Bth)/\nu \geq \mathcal F^{-1}_Q[\rhoth]/\nu.
\ee
The second inequality is referred to as the quantum (multiparameter) Cramér-Rao bound (QCRB), which defines the
{\it quantum Fisher information matrix} (QFIM)
\be
[\Qfishrho{\rhoth}]_{jk} := \tr[\rhoth (L_j L_k + L_k L_j)/2] ,
\ee
where the $L_i$ are the symmetric logarithmic derivatives (SLDs) relative to each of the phases, i.e., they satisfy
\be
\partial_{j} \rhoth = \frac 1 2 (L_j \rhoth + \rhoth L_j) ,
\ee
where we used the simplified notation $\partial_{j}=\partial_{\theta_j}$. Note that a more general version, termed Holevo-Cram\'er-Rao bound can be also found, which is generally tighter than \cref{eq:qCRboundmulti}~\cite{yuen1973multiple,belavkin2004generalized,holevo2011probabilistic,AlbarelliFrielDatta2019,Sidhuetal2021}. When the probe state is pure, i.e., $\rhoth=\ketbra{\Psi_\Bth}$, the QFIM is proportional to the covariance matrix of the vector ${\bf \tilde{H}}(\vec \theta)$ of generators~\cite{Baumgratz_Datta2016,liu2020quantum}
\be
\tilde H_j(\Bth):=iU^{\dagger}(\Bth)\partial_{j} U(\Bth) = \int_0^1 e^{is{\bf H} \cdot \Bth} H_j e^{-is{\bf H} \cdot \Bth} {\rm d}s , \label{observable}
\ee
calculated on the initial state $\ket{\Psi_0}$.
Namely, we have 
\be\label{qfim}
\Qfishrho{\Psi_\Bth}=4\Covv{{\bf \tilde{H}}(\Bth)}{\Psi_0} ,
\ee
where the (symmetric) covariance matrix of a vector of operators ${\vec A}=(A_1,\dots, A_K)$ on a quantum state $\varrho$ is defined as
\be
[\Covv{{\vec A}}{\varrho}]_{jk} := \tfrac 1 2 \ex{A_j A_k + A_k A_j}_\varrho - \ex{A_j}_\varrho \ex{A_k}_\varrho .
\ee
From the matrix inequality in \cref{eq:qCRboundmulti} one can derive scalar figures of merit for the performance of the estimator from any positive semidefinite weight matrix $P$ via $\tr[P \Qfishrho{\rhoth}]$. A common choice is just to take $P=\id$ and thus obtain
\be\label{eq:trCovvstrF}
\sum_i \var{\hat \theta_i} \geq \frac 1 \nu \tr[\mathcal F_Q^{-1}[\rhoth]] .
\ee
In some cases where the group structure of the unitaries is relevant, e.g., in the $SU(2)$ case that we are going to discuss, another approach is
to consider the metric as a group-covariant weight matrix~\cite{GoldbergSanchezSotoFerrettiPRL2021}.

Note also that the matrix inequality in 
\cref{eq:qCRboundmulti} is saturated if and only if all possible inequalities for the scalar figures of merit are saturated. Thus, the saturation of one single scalar condition is in general a necessary but not sufficient requisite for the saturation of the matrix inequality.

In the multiparameter case the quantum Cram\'er-Rao bound in \cref{eq:qCRboundmulti} is in general harder to saturate than in the single-parameter case, even in the asymptotic limit, due to the fact that the SLDs associated to the different parameters might not commute. In fact, a simple condition for saturation that is in general only sufficient but not necessary is given by $[L_j,L_k] = 0$ for all $j$ and $k$, which implies that all the SLDs share a common eigenbasis and thus the optimal measurement is simply given by a projective measurement onto that basis~\cite{Matsumoto_2002,Ragy_2016,Liu_2016}. 
A simple necessary and sufficient condition can be instead given for pure states $\rhoth=\ketbra{\Psi_\Bth}$ and reads
\be\label{eq:suffcondsatMCRbound}
\bra{\Psi_0} [\tilde H_j(\Bth), \tilde H_k(\Bth)] \ket{\Psi_0} =0 \qquad \text{for all } j,k .
\ee
Whenever the saturability condition is satisfied, concrete conditions for POVMs to saturate the QFIM bound have been derived in \cite{Baumgratz_Datta2016}, while in \cite{pezzeetalPRL2017multi} it was shown how measurements that saturate the QCRB for a pure probe state can be constructed by means of projectors into the probe state plus a set of projectors into its orthogonal subspace.

\subsection{Method of moments estimator and accuracy}

When the Cram\'er-Rao bound is saturated, there are well-known estimators that saturate it in the limit of large $\nu$, one of which is given by the maximum-likelihood estimator $\hat \theta = \mathrm{argmax}_\theta \, {\rm p}(q | \theta)$, which, however, requires sampling extensively the likelihood function ${\rm p}(q | \theta)$ (seen as a function of $\theta$). A common alternative, which only requires determining lower-order moments of ${\rm p}(q|\theta)$, is the so-called {\it method of moments} (MoM)~\cite{pezzerev18,gessnerNatComm20,Kay1993FSSP_Estimation}, which we are going to consider in the following. Let us illustrate it first in the single parameter scenario. One performs projective measurements $\Pi_q=\ketbra{q}$ in the eigenbasis of an observable $M=\sum_q m_q \, \ketbra{q}$ and uses as an estimator for $\theta$ the expectation value 
\be
\ex{M}_{\varrho_{\theta}} = \sum_q m_q \tr[\Pi_q \rho_\theta] = \sum_q m_q \tr[U^\dagger(\theta) \Pi_q U(\theta) \rho_0] .
\ee
More precisely, the estimator in this case will be given by the value $\thetaMoM$ such that 
\be
\mu(\thetaMoM):=\bar M_\nu = \ex{M}_{\varrho_{\theta}}  ,
\ee
where $\bar M_\nu = \tfrac 1 \nu \sum_{i=1}^\nu m_{q_i}$ is the measured average, calculated from $\nu \gg 1$ measurement outcomes $\{ q_i\}_{i=1}^\nu$~\cite{pezze2014quantum,pezzerev18,gessnerNatComm20}.
In other words, this estimator is given by
\be
\thetaMoM := \mu^{-1}(\bar M_\nu) ,
\ee
where $\mu^{-1}(\cdot)$ is the inverse of the function given above. This estimator becomes unbiased in the large $\nu$ limit and its accuracy can be calculated from the error-propagation formula, and, to the first order in $1/\nu$, is given by:
\be
(\Delta \thetaMoM)^2 = \frac 1 \nu \frac{(\Delta M)_{\varrho_\theta} ^2}{|\partial_\theta \ex{M}_{\varrho_\theta} |^2} = 
\frac 1 \nu \frac{(\Delta M)_{\varrho_\theta} ^2}{|\ex{[M,\tilde H]}_{\varrho_\theta} |^2} . 
\ee
This formula (once again, valid in the limit of $\nu \gg 1$) is calculated from a Taylor expansion of the function $\mu^{-1}(x)$ in the neighborhood of $\langle M\rangle_{\varrho_\theta}$. More details about this method are provided in \cref{app:MoMdetails}.

This method of moments can be generalized to the multiparameter scenario~\cite{gessnerNatComm20}. In this case, one calculates the expectation value of a set of observables ${\vec M} = (M_1,\dots,M_K)$ and again estimates $\vec \theta$ from the corresponding sample means after $\nu \gg 1$ rounds: $\mean{\vec M}_{\varrho_{\vec \theta}} = \bar{\vec M}_\nu$. The accuracy can be again calculated via a Taylor series expansion of the multivariate estimator around $\mean{\vec M}_{\varrho_{\vec \theta}} $, but it also becomes important to distinguish whether the observables $\vec M$ commute with each other or not. If they all commute, their joint expectation value can be obtained from a single measurement in the common eigenbasis $\ketbra{q}$ and the accuracy becomes (at first order in $1/\nu$):
\be
\begin{aligned}\label{eq:boundprecisionfrommoment}
\COV(\Th) &= \mathcal M_{\rhoth}^{-1}[{\tilde{\vec H}}, {\vec M}]/\nu \\
&= \tfrac 1 \nu \Omega_{\rhoth}^{-1}[{\tilde{\vec H}}, {\vec M}] \Covv{{\vec M}}{\rhoth} (\Omega_{\rhoth}^{-1}[{\tilde{\vec H}}, {\vec M}] )^T ,
\end{aligned}
\ee
where $\Covv{{\vec M}}{\rhoth}$ is the covariance matrix of the set of observables ${\vec M}$ and 
\be
[\Omega_{\rhoth}[{\tilde{\vec H}}, {\vec M}]]_{kl} = - i \ex{[M_k,\tilde H_l]}_{\rhoth}
\ee
is the commutation matrix, both of them calculated on the evolved probe $\rhoth$. When the observables $\vec M$ do not commute, their expectation values have to be obtained either from separated subsets of the samples, or by performing a POVM that approximates a joint measurement of all the $M_k$, which, however, has generally increased uncertainty~\cite{busch2016quantum}. See \cref{app:multiMoM} for further details. In \cite{gessnerNatComm20}, it has also been investigated how to look for bounds on the precision reached by the method of moments, when fixing the set of available observables $\vec A$, which includes the generators of the dynamics, as well as additional observables that are considered as measurements, which is what we are also going to do in what follows. For finding this overall bound, one defines the moment matrix of the available observables as
\begin{equation}
\mathcal{M}_\varrho[\vec A]
:=\Omega^T_\varrho(\vec A) \;\Cov^{-1}_\varrho(\vec A)\;\Omega_\varrho(\vec A) ,
\label{eq:accessible_moment_matrix}
\end{equation} 
where $\bigl[\Omega_\varrho(\vec A)\bigr]_{kl}:=-\mathrm{i}\,\langle[A_k,A_l]\rangle_\varrho$ is the matrix of expectation values of commutators. The following upper-bound then holds:
\begin{equation}
\mathcal{M}_\varrho\bigl[\tilde{\vec H}(\vec\theta),\vec M\bigr]
\;\le\;
R(\vec\theta)\,\mathcal{M}_\varrho[\vec A]\,R(\vec\theta)^T,
\label{eq:moment_matrix_upper_bound} 
\end{equation}
where $R$ is the matrix such that $\tilde{\vec H} = R \vec A$. See \cref{app:mom_optimal_bound} for more details.

\section{Intermezzo: Estimation of Heisenberg-Weyl unitaries}\label{sec:HWintermezzo}

As a preliminary discussion on the idea we want to employ, let us consider now arguably the simplest scenario of estimation of multiple phases from non-commuting generators, which is the case in which the generators form the Heisenberg-Weyl algebra:
\be
[x,p]= i\id .
\ee
This essentially means that we consider a single-mode system and want to estimate a two-parameter phase-space displacement, i.e.:
\be
U(\vec r)=e^{-i(r_1 x + r_2 p)} .
\ee
Let us imagine that we want to estimate $\vec r = (r_1 , r_2)$ by measuring the quadratures $\vec X = (x,p)$ and using the method of moments. For these observables, the expectation value of the commutator is actually constant for all states and thus the commutator matrix is given by the canonical symplectic form
\be
\Omega_\varrho[\vec X] = \begin{pmatrix}
    0 & 1 \\
    -1 & 0
\end{pmatrix} := \Omega ,
\ee
independent of $\varrho$.
For a given probe state $\varrho$, we then consider the moment matrix of the quadrature vector $\vec X$ which is given by
\be
\mathcal M_\varrho[{\vec X}] =\Omega^T \; \Gamma^{-1}_{\varrho}[\vec X] \; \Omega ,
\ee
which for Gaussian states $\varrho_G$ actually coincides with the QFIM~\cite{gessnerNatComm20}. In particular, for pure Gaussian probe states $\ket{\Psi_{\vec r}} = U(\vec r)\ket{\Psi_{\vec 0}}$ we can use \cref{qfim} and also obtain directly
\be
\Qfishrho{\Psi_{\vec r}} = 4 \Covv{\tilde{\vec X}}{\Psi_{\vec 0}} .
\ee
For simplicity, let us consider the case $\vec r \simeq \vec 0$, which implies $\tilde{\vec X} = \vec X$. 
Note that in this case the saturability condition in \cref{eq:suffcondsatMCRbound} is not met for any pure state, since the commutator between the generators is 
a nonzero constant. 

On the other hand, because the commutator between the generators does not vanish, we can take the vector of available observables to coincide with the generators, i.e:
\be
\vec H = \vec X \implies R= \id .
\ee
Correspondingly, we can see that a vector of optimal measurements is given by 
\be
\vec M = \Omega \; \Cov^{-1}_{\varrho}[\vec X] \; \vec X ,
\ee
which follows from the general argument discussed in \cref{app:mom_optimal_bound} (see \cref{eq:Mopt_general} with $G=R=\id$).

Now it just remains to analyze the probe state. As a first example, let us consider the vacuum state $\ket{0}$.
For such a state the covariance matrix is just proportional to the identity and we have
\be
\mathcal M[{\vec X}] =  \Omega^T \; \Cov^{-1}_0[\vec X] \; \Omega = 2\; \id ,
\ee
while the optimal observables whose expectation values have to be measured are
\be\label{eq:vecMforHW}
\vec M = \Omega \vec X =  (p , -x) ,
\ee
which is essentially the same as the vector $\vec X$ itself.

The same reasoning would hold for all coherent states, defined by
\be
\frac 1{\sqrt 2}(x + ip) \ket{\alpha} = \alpha \ket{\alpha} ,
\ee
where $\alpha$ is a complex number.

Let us now instead consider a squeezed state, defined by 
\be
\ket{\xi} = e^{\tfrac 1 2 \xi (a^2 - (a^\dagger)^2)} \ket{0} ,
\ee
where $\xi$ is a real parameter which gives the squeezing factor.

In this case we have $\Covv{\vec X}{\xi} = \tfrac 1 2 {\rm diag}(e^{-2\xi},e^{2\xi})$ and consequently 
\be
\mathcal M_\xi [{\vec X}] = 2  {\rm diag}(e^{-2\xi},e^{2\xi}) ,
\ee
while the optimal observables can be again taken to be those in \cref{eq:vecMforHW}.

The question then arises how to directly estimate from a single measurement run the expectation values of both $x$ and $p$ quadratures. For that, in this case one can employ a heterodyne measurement, which is essentially a POVM into the (non-orthogonal) coherent states projectors $\ketbra{\alpha}$.
Formally, we have that the resolution of identity is given by
\be\label{eq:eterodynePOVM}
\id = \frac 1 {\pi} \int_{\mathbb C} {\rm d}^2 \alpha \  \ketbra{\alpha} .
\ee
Such a measurement gives as an outcome a phase-space point labelled by $\alpha \in \mathbb C$ and is such that its real and imaginary parts provide the values of $x$ and $p$. Thus, by estimating the expectation value of $\alpha$, we get an estimate of both the expectation values of $\ex{x}$ and $\ex{p}$ through its real and imaginary parts respectively. This essentially amounts to evaluate the Husimi function of the quantum state, namely:
\be
\mathcal Q_\varrho(\alpha) = \frac 1 \pi \bra{\alpha} \varrho \ket{\alpha} .
\ee

Note also that for the MoM estimator we do not need the full statistics of $\alpha$, but just its first order moments, i.e., the expectation value $\ex{\alpha}$, which may be obtained also from such an imperfect measurement. Still, in order to evaluate the accuracy one needs to calculate the moment matrix, which in this case is essentially given by the covariance matrix $\Gamma_\varrho[\vec X]$. This can be also reconstructed from heterodyne measurements (i.e., from a partial reconstruction of the Husimi function) or other imperfect joint measurements of the quadratures, but with additional noise coming from the non-commutativity of the quadratures and the non-existence of a perfect joint measurement.

We remark that in this case the QCRB is not saturated, and this example just provides an illustration of how a multiparameter estimation can be performed using the method of moments, also
for the case of two noncommuting generators.

\section{Multiparameter $SU(2)$ estimation}\label{sec:SU2multi}

\begin{figure}[h]
\centering
\includegraphics[width=0.97\columnwidth]{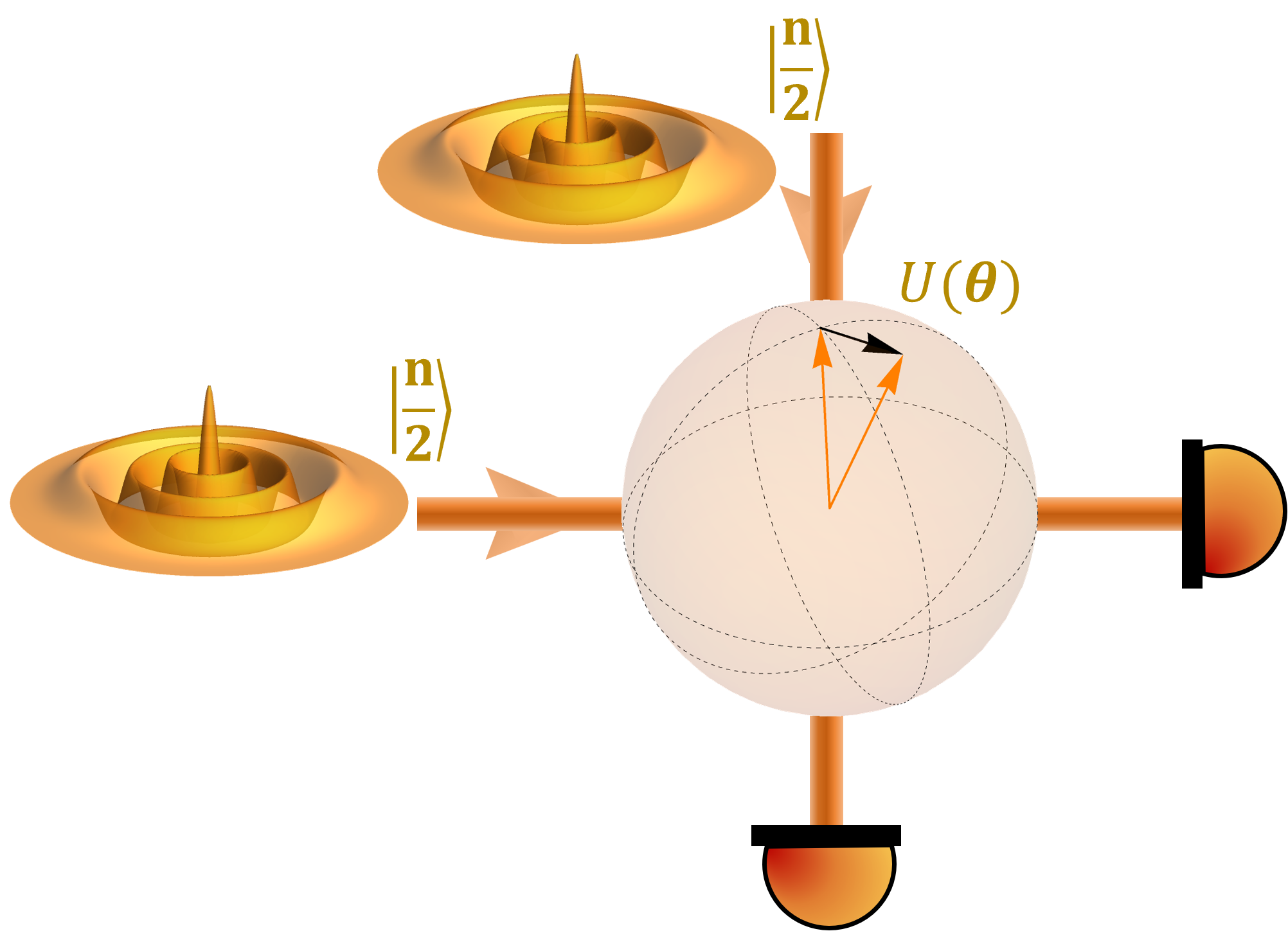} 
\caption{Three-parameter $SU(2)$ estimation. An input state $\probez$ undergoes a $SU(2)$ rotation, evolving into the state $U(\Bth)\probez$, and then detected at two output ports. The input is a two-mode state of the form \cref{eq:probeXYZ}, like the twin Fock state obtained for $k=n / 2$.
}
\label{figPara3SU2}
\end{figure}

Let us now consider an estimation of a $SU(2)$ unitary parametrized by three phases, as depicted in \cref{figPara3SU2}. In the literature this problem has been considered in a number of works, also from the point of view of abstract estimation of unitaries, without referring to a specific parametrization~\cite{GoldbergSanchezSotoFerrettiPRL2021}, and with general group-invariant figures of merit \cite{Martin2020optimaldetectionof,AcinetalPRA2001,ChiribellaDarianoPerinottiSacchi,ChiribellaDarianoSacchiPRA05,GoldbergSanchezSotoFerrettiPRL2021}. Here, following the previous discussion and other typical approaches in quantum metrology~\cite{Vaneph_2013,Baumgratz_Datta2016,GoldbergJames2018,Z_Goldberg_2021,Gorecki2022,gessnerNatComm20}, we are interested in the estimation of a three-dimensional vector of phases $\Bth = (\theta_x,\theta_y,\theta_z)$ related to a given parametrization of the unitary, and in minimizing the corresponding variances, close to a reference point. That is, we focus on the common approach of {\it local} estimation, where however the reference point, as well as the parametrization can be in principle arbitrary. Applying the Cram\'er-Rao bound and using the convexity of the QFIM, we will thus look for pure probe states that maximize the covariances of the observables in \cref{observable}, while ensuring the saturability condition in \cref{eq:suffcondsatMCRbound}. These provide the best possible resource, at least in an asymptotic regime of a large number of independent repetitions. In this case of an $SU(2)$ unitary, it is easy to see that the generators defined from \cref{observable} form an $su(2)$ algebra (see also \cref{app:detailsFullSU2}).

For illustrating the known results with more simplicity, we consider here small values of the phases $(\theta_x,\theta_y,\theta_z)\simeq (0,0,0)$ such that these observables can be simply set to be the three canonical spin components. In \cref{app:detailsFullSU2} we discuss the more general case of arbitrary phases, as well as different parametrizations (still in a local point estimation scenario, where the QCRB is meaningful), and observe how in the case of $SU(2)$ estimations the general case can be worked out explicitly, due to the compactness of the group. For infinitesimal phase imprintings, the probe state is given by\footnote{Note that, even if globally an $SU(2)$ unitary can have multiple parametrizations, in the case that we consider of infinitesimal unitaries this form is valid independently of the chosen parametrization.} 
\be
\ket{\Psi_\Bth} \simeq (\id - i \vec \theta \cdot \vec J) \probez ,
\ee
where $\vec J =(J_x , J_y , J_z)$ are a basis of the $su(2)$ algebra, i.e., three orthogonal reference spin axes, which in our context are given by the Schwinger representation
\be
\begin{aligned}
    J_x&=\frac{1}{2}(a^{\dagger}_1a_2+a_1a^{\dagger}_2) \\ 
    J_y&=\frac{1}{2i}(a^{\dagger}_1a_2-a_1a^{\dagger}_2) \\  
    J_z&=\frac{1}{2}(a^{\dagger}_1a_1 - a^{\dagger}_2a_2) , \label{spinop}
\end{aligned}
\ee
with two bosonic modes satisfying canonical commutation relations $[a_i , a_j^\dagger]=\delta_{ij}\id$. Thus, in this infinitesimal regime we have $\tilde{\vec H} = \vec J$ and the QFIM is given by
\be
\Qfishrho{\Psi_{\vec 0}} = 4 \Covv{{\vec J}}{\Psi_{\vec 0}} .
\ee
Note that eigenstates of spin components will always have a singular QFIM and in that case the QCRB is meaningful only in the two-dimensional subspace given by the two orthogonal spin components.

\subsection{Probe states reaching Heisenberg scaling for three parameters}

To analyze potentially good candidate (pure) probe states, we first consider sectors with fixed number of particles $n$ and take this as the resource with respect to which we investigate the precision scaling. This is a common approach, which relates to the energy or intensity of the probe. Moreover, in this context where the evolution preserves the particle number, components of the states in each fixed-$n$ sector will be invariant during the evolution, and the question also arises on whether the measurement can detect superpositions between different sectors or not. Later in \cref{sec:indefN} we will discuss more extensively probe states with a non-fixed particle number. In this scenario, it is well-known that the best single-parameter probe state is the NOON state~\cite{TothApellaniz2014,pezzerev18,Demkowicz_Dobrza_ski_2015} (see also \cref{app:recapsinglesu2}), which has a non-singular QFIM and is an eigenstate of $J_z^2$. On a deeper inspection one can then see that, while it provides Heisenberg scaling for estimating one phase, say $\theta_z$, its QCRB for the other two phases scales as the SQL. Another well-known good candidate state is the twin-Fock state, which is also an eigenstate of $J_z$ and $J^2_z$ at the same time and thus, while it is useful only for the estimation of two phases, $\theta_x$ and $\theta_y$, it allows achieving Heisenberg scaling in both of them. In general, eigenstates of $J_z^2$ with $\langle\vec{J}\rangle=0$ have a diagonal QFI matrix and the QCRB is saturated in their case, since \cref{eq:suffcondsatMCRbound} is satisfied.

This is also an indication that eigenstates of $J_z^2$ can be considered as good candidate probe states in the multiparameter case. In fact, other eigenstates of $J_z^2$ present sort of intermediate scalings for the three phases and, in particular one can find some that allow Heisenberg scaling for all three directions. Specifically, these have the form
\be\label{eq:probeXYZ}
\ket{\Psi_k}:=\frac1{\sqrt 2}\left(\ket{n-k,k} + \ket{k,n-k} \right),
\ee
with $k= \left\lfloor \frac{n-\sqrt{\frac{n(n+2)}{3}}}{2} \right\rceil := k_{\rm opt}$ and $\lfloor x \rceil$ denoting the closest integer to $x$~\footnote{Note that here, and similarly later in \cref{eq:su11states} the $1/\sqrt 2$ normalization of the state should be modified in order to obtain the twin Fock state as a special case in the class.}.

The QFIM of states of the form \eqref{eq:probeXYZ} is given by
\be 
\Qfishrho{\Psi_k}= {\rm diag}\left(F_\parallel,F_\parallel, F_\perp \right) ,
\ee
with $F_\parallel = 2k(n-k)+n$ and $F_\perp = (n-2k)^2$, which reaches Heisenberg scaling for all parameters already for $k\sim O(n)$.

In the case of $k= k_{\rm opt}$ the state becomes close to those called two-anti-coherent in literature~\cite{Bouchard:17,ChryssomalakosCoronado2017,GoldbergJames2018,Martin2020optimaldetectionof,Z_Goldberg_2021,Serrano_Ens_stiga_2025}  (see also \cite{piotrak2023perfect} for a similar approach), which are states that have equal value of second moments of the angular momentum components for all possible directions $\hat{\rm n}$, i.e., $\ex{J^2_{\hat{\rm n}}}=j(j+1)/3$ is satisfied for all directions $\hat{\rm n}$~\footnote{Note that anti-coherent states are defined for any order $M$. Those are states with equal spin moments of order $M$ in any direction~\cite{Bouchard:17,ChryssomalakosCoronado2017,GoldbergJames2018,Martin2020optimaldetectionof,Z_Goldberg_2021}. In this case we are discussing the case $M=2$.}. Due to this similarity, we can refer to the state with $k= k_{\rm opt}$ as {\it quasi}-two-anticoherent.

In \cite{Z_Goldberg_2021} two-anticoherent states were analyzed in detail for the task of estimating three rotation angles from a geometrical point of view and also optimal measurements were studied, following the approach of \cite{pezzeetalPRL2017multi} that generates an optimal PVM from the probe state and its orthogonal subspace. Furthermore, they also analyzed the case of a POVM of the form analogous to that in \cref{eq:eterodynePOVM}, which relates to the Husimi function in this $SU(2)$ context. Specifically, approaching the analysis from a geometrical perspective the authors argued for a reasonable subset of projectors onto spin-coherent states that need to be measured in order to obtain a good scaling of the classical Fisher information matrix for all three parameters.

Thus, we can see that the states from the class \cref{eq:probeXYZ} that are close to be two-anticoherent would also allow reaching Heisenberg scaling for a full three-parameter $SU(2)$ encoding in an analogous fashion, except that they are simpler, which also means that the optimal PVM that allows to saturate the QCRB is correspondingly a bit simpler as well, as it involves the projector into the probe state itself, plus a set of orthogonal projectors. Similarly, we would expect that the set of projectors onto coherent spin states that need to be measured to approach the saturation of the QCRB bound would be similar. In the next subsection, we will discuss an approach that is partly analogous, but focused on the method of moment estimator, and we restrict to an estimation that is based on just the reconstruction of the first two moments of the generators. In that case, as we will see, it is not possible to saturate the QCRB for the full three-parameter estimation.

As a further comparison with two-anticoherent states, we mention that the three-parameter $SU(2)$ estimation can also be seen as a rotation sensing scenario, where the rotation direction is not known, i.e, a unitary encoding of the form $U_{_{\hat{\vec{\rm n}}}}(\theta)=\exp(-i\theta J_{\hat{\vec{\rm n}}})$, where $\hat{\vec{\rm n}}$ is an arbitrary direction. With such a viewpoint, a natural benchmark figure of merit is the average performance over all directions~\cite{Z_Goldberg_2021,HervasetalPRL2025}, which is lower bounded from the QCRB as:
\be
\overline {\Delta^2\hat \theta} \geq \frac1{4\pi}\int {\rm d} \hat{\vec{\rm n}} \frac1{F_Q[\ket{\Psi_{\theta}}, J_{\hat{\vec{\rm n}}}]} .
\ee
In this respect, it is known that two-anticoherent states are optimal~\cite{Z_Goldberg_2021}, and would remain optimal probes also when the first correction to the QCRB with the maximum-likelihood estimator is considered~\cite{HervasetalPRL2025}.

Once again, in analogy to this reasoning, considering as probe states those in \cref{eq:probeXYZ} we obtain
\begin{equation}
    \bar F_Q[\ket{\Psi_k} ,J_{\hat{\vec{\rm n}}}]:=\frac1{4\pi}\int {\rm d} \vec n \ F_Q[\ket{\Psi_k}, J_{\hat{\vec{\rm n}}}]=\frac13n(n+2),
\end{equation}
which is exactly the same as the two-anticoherent states~\cite{Z_Goldberg_2021}. In terms of averaged accuracy we get, considering for concreteness only the case $F_\perp = F_\parallel + \delta \simeq F_\parallel$ (i.e., with $|\delta| \ll 1$):
\begin{equation}
    \begin{aligned}
        \overline {\Delta^2\hat \theta}&\geq \frac1{4\pi}\int {\rm d} \vec n \ \frac{1}{F_Q[\ket{\Psi_k}, J_{\hat{\vec{\rm n}}}]} \\
        &= \frac 1 {\sqrt{F_\parallel(F_\perp-F_\parallel)}}{\tan^{-1}\sqrt{\frac{F_\perp-F_\parallel}{F_\parallel}}} \\
        &= \frac 1 {F_\parallel} - \frac{\delta}{3 F_\parallel^2} + O(\delta^2) ,
    \end{aligned}
\end{equation}
which as expected contains a $O(\delta)$ correction with respect to the case of two-anticoherent states, where the result is $\overline {\Delta^2\theta}_{\rm anti}=\frac{3}{n(n+2)}$~\cite{Z_Goldberg_2021}. (More broadly, a similar expression can be obtained for $F_\perp > F_\parallel$).

To conclude, we point out that other states that achieve Heisenberg scaling for the three phases were studied in Refs.~\cite{AlbarelliFrielDatta2019,Baumgratz_Datta2016}, where they also considered the case of arbitrary phases and the optimal measurements. More precisely, they studied a probe state which is a superposition of three NOON states in the three orthogonal directions and the optimal measurements that are more realistically implementable are composed of projectors of the form $(\id \pm \sigma_k^{\otimes n})/2$ with $k = \{x,y,z\}$.

\subsection{Estimation via multiparameter method of moments}

As we have seen, states are known that can be useful to estimate three phases with Heisenberg scaling, and also measurements saturating such bounds can be worked out. However, these schemes can be quite challenging in some setups. Also, from a purely theoretical perspective, it is interesting
to look for alternative protocols and, at the same time, uncover other aspects that might also be related to the group-structure of the unitaries to be estimated.
Here, we have already discussed that states of the form \eqref{eq:probeXYZ} can be a viable alternative to those discussed in literature for the potential achievement of Heisenberg scaling, which in some sense is also constructed from the algebraic structure forming the unitary group. In fact, we will observe a somewhat similar structure later for the case of $SU(1,1)$ unitaries. 

On top of that, we can also try to observe whether a pragmatic estimation scheme, namely the method of moments implemented with reasonable observables, also allows the saturation of Heisenberg scaling, or alternatively what is the best scaling achievable in that case. In particular, we ask the question: What is the best achievable precision (scaling) when only the first two moments of the generators are available? Clearly, more in general this question can be asked for different sets of observables, which may be relevant in different experimental contexts. The case we consider here can be seen as being of relatively general relevance, as it involves the generators of the dynamics themselves. Moreover, the first two moments are what is most typically estimated in actual experiments, especially those where the particle number is conserved, e.g., atomic ensembles~\cite{pezzerev18}, while larger and larger order moments become quickly very tedious to consider and also more difficult to reconstruct in practical circumstances. 

Thus, overall for the estimation of an $SU(2)$ unitary we consider the following vector of observables:
\be\label{eq:vecAsu2}
\vec A_{\mathfrak{su}(2)} = (\vec J, \vec Q ) ,
\ee
where 
\be
\vec Q = (J_x^2,J_y^2,J_z^2, \{J_x , J_y\} , \{J_x , J_z\} , \{ J_y , J_z \})
\ee
is the vector of quadratic products of spin operators.

It is also useful to express these observables in a two-index tensor notation from the quadratic products
\be
A_{kl} = J_k J_l ,
\ee
such that by taking the symmetric and anti-symmetric part we can write
\be\label{eq:JandQmatrices}
\begin{aligned}
    J_i &= \sum_{kl} \epsilon_{ikl} A_{kl} , \\
    Q_{kl} &= A_{kl} + A_{lk},
\end{aligned}
\ee
with $\epsilon_{ikl}$ being the usual Levi-Civita symbol and we see that basically $\vec A_{\mathfrak{su}(2)}$ is just a vectorized form of the operator matrix $A$.

Furthermore, the expectation value of the symmetric operator matrix $Q$ on a quantum state gives the correlation matrix of the spin operators. Thus, estimating the expectation values of the vector of observables $\vec A_{\mathfrak{su}(2)}$ essentially amounts to estimating the average vector $\ex{\vec J}= (\ex{J_x}, \ex{J_y} , \ex{J_z})$ and the covariance matrix $\Cov_{\Psi_{\Bth}}[{\vec J}]$ on the probe state $\ket{\Psi_{\Bth}}$. Here, we work in the Heisenberg picture, in which the probe state remains fixed and the elements of our vector of observables $\vec A$ are transformed as
\be
A_i(\vec \theta) = U^\dagger(\vec \theta) A_i U(\vec \theta) .
\ee
Group-theoretically speaking, these observables transform under the adjoint representation of $SU(2)$, and in particular the vectors $\vec J$ and $\vec Q$ are transformed as
\be
\begin{aligned}
    \vec J(\vec \theta) &= U^\dagger(\vec \theta) \vec J U(\vec \theta)  = \mathcal O_{\bf 3} (\vec \theta) \, \vec J  , \\ 
    \vec Q(\vec \theta) &= U^\dagger(\vec \theta) \vec Q(\vec \theta) U(\vec \theta)  = \mathcal O_{\bf 6} (\vec \theta) \, \vec Q ,
\end{aligned}
\ee
where $\mathcal O_{\bf 3}(\vec \theta)$ and $\mathcal O_{\bf 6}(\vec \theta)$ are $3\times 3$ and $6 \times 6$ orthogonal matrices respectively which are obtained from the adjoint representation of $SU(2)$ (see \cref{app:detailsFullSU2} for details). The matrix $\mathcal O_{\bf 3}$ gives basically a three-dimensional rotation of the spin axes, while the six-dimensional matrix is constructed from vectorizing the quadratic relation
\be
Q(\vec \theta) = \mathcal O_{\bf 3} \, Q \, \mathcal O^T_{\bf 3} , 
\ee
where $Q$ is the symmetric operator matrix given in \cref{eq:JandQmatrices}, which is transformed quadratically with the three-dimensional orthogonal rotation of the spin axes. Further details are provided in \cref{app:detailsFullSU2}. 

Let us now consider the moment matrix $\mathcal M_{\Psi_\Bth}[\vec A_{\mathfrak{su}(2)}]$ coming from \cref{eq:boundprecisionfrommoment} for the vector of observables as in \cref{eq:vecAsu2}. Here, the probe states are the eigenstates $\ket{\Psi_k}$ of $J_z^2$ as in \cref{eq:probeXYZ} for all possible values of $k$. As we mentioned, this class contains states that can be potentially good resources for multiparameter $SU(2)$ estimations. Then, we can calculate what the optimal observables $\vec M$ are among the vector $\vec A_{\mathfrak{su}(2)}$, which can be done following the methods described in \cref{app:MoMdetails} (cf. \cref{eq:Mopt_general}). In summary, we have 
\be
\begin{aligned}
    \mathcal M_{\Psi_\Bth}[{\vec H}, {\vec M}] &\leq  R \mathcal M_{\Psi_\Bth}[{\vec A}] R^T \\
    &= R \, \Omega^T_{\Psi_{\vec \theta}}[{\vec A}] \, \Gamma^{-1}_{\Psi_{\vec \theta}}[{\vec A}] \, \Omega_{\Psi_{\vec \theta}}[{\vec A}] R^T ,
\end{aligned}
\ee
where $R$ is the matrix such that
\be
\vec J = R \vec A .
\ee
This way, we get from \cref{eq:boundprecisionfrommoment} an upper bound to the precision matrix given by
\be
\begin{aligned}
    \COV^{-1}(\Th) &\leq \nu R \, \Omega^T_{\Psi_{\vec \theta}}[{\vec A}] \, \Gamma^{-1}_{\Psi_{\vec \theta}}[{\vec A}] \, \Omega_{\Psi_{\vec \theta}}[{\vec A}] R^T .
\end{aligned}
\ee
Note that the saturability of this bound is not trivial in general, and eventually can be reached only in the asymptotic limit $\nu \gg 1$. Note also that this relation assumes that the moment matrix is invertible. However, if there are some zero eigenvalues of $\mathcal M_{\Psi_\Bth}(\vec A)$ we should discard the corresponding observable from the vector $\vec A$ since it will simply not contribute at all to the estimation.

Now let us consider probe states of the form \cref{eq:probeXYZ}, and again for simplicity an infinitesimal estimation with $\vec \theta\simeq \vec 0$. Because all of the states in our family $\ket{\Psi_k}$ are eigenstates of $J_z^2$ (the corresponding eigenvalue being $\frac14 (n-2k)^2$ for a given value of $k$) it turns out that the expectation values of $J_z$ and $J_z^2$ are insensitive to the (infinitesimal) rotation we want to estimate. Thus, first of all with such states it is impossible to estimate $\theta_z$ by measuring expectation values of $\vec A_{\mathfrak{su}(2)}$; secondly, the observables $J_z$ and $J_z^2$ can be eliminated from $\vec A_{\mathfrak{su}(2)}$ entirely since they are not useful also for the estimation of $\theta_x$ and $\theta_y$.
Consequently, for the states in \cref{eq:probeXYZ} we can consider the reduced vector of observables 
\be\label{eq:vecAprimesu2}
\vec A_{\mathfrak{su}(2)}^\prime=(J_x,J_y,\{J_x,J_z\},\{J_y,J_z\}) ,
\ee
and the reduced problem of estimating just $\vec{\theta}^\prime =(\theta_x,\theta_y)$. Moreover, as we already briefly observed before, for the two-parameter estimation problem it is actually the twin-Fock state that gives the best precision scaling (which corresponds to the state with $k=n/2$ in \cref{eq:probeXYZ}). What we observe here is that Heisenberg-scaling precision can in principle be maintained also via the method of moments with the observable set in \cref{eq:vecAprimesu2}. Note, however, that in practice it might still be challenging to make a suitable (potentially imperfect) reconstruction of the moments needed for the estimation, as the observables do not commute.

Making a more precise statement, we find that none of the states of the form \cref{eq:probeXYZ} have a moment matrix $\mathcal M_{\Psi_\Bth}[{\vec A}]$ that allows to reach Heisenberg scaling for all three parameters $\Bth=(\theta_x ,\theta_y, \theta_z)$. We find that only the twin-Fock state (or states close to it, e.g., with $k=n/2 \pm 1$) allow for Heisenberg-scaling precision and only for the reduced problem of estimating $\vec{\theta}^\prime$. In particular, for the twin-Fock state we get the following precision matrix (for small $\Bth$, but analogous statement can be made for arbitrary $\Bth$ in this case, see \cref{app:detailsFullSU2}):
\be 
\lim_{\Bth \rightarrow (0,0,0)} \COV^{-1}(\hat{\vec{\theta}}) 
\leq \nu \ {\rm diag}\left(\frac{n (n+2)}{2} , \frac{n (n+2)}{2} , 0 \right) ,
\ee 
and thus we see that in this case the theoretical Heisenberg scaling of the QFIM could be in principle saturated via method of moments.

Furthermore, we can also obtain the optimal observables from which to evaluate the expectation values, which are (cf. \cref{eq:Mopt_general})
\be 
\vec M^\prime=(\{J_x,J_z\},\{J_y,J_z\})^T ,
\ee 
i.e., essentially one needs to estimate cross-covariances between the spin operators. Note that in \cite{gessnerNatComm20} it was also discussed how to saturate the QCRB (i.e., achieve simultaneous Heisenberg scaling for $\theta_x$ and $\theta_y$) for the twin-Fock state via the method of moments. It was pointed out that a set of two commuting observables can be found, which are $A_x = J_x \ketbra{\Psi_{n/2}}J_x$ and $A_y = J_y \ketbra{\Psi_{n/2}}J_y$. But they involve the projector onto the twin-Fock state itself.

We conclude this discussion by briefly commenting on another candidate state that was proposed to achieve the Heisenberg limit for the three parameters, namely the state that is a superposition of three NOON states relative to three orthogonal directions~\cite{AlbarelliFrielDatta2019,Baumgratz_Datta2016}. For this, one can observe that the method of moments involving only quadratic spin observables does not work, essentially because the moment matrix would be entirely zero for that state. In fact, as we already mentioned, optimal measurements that achieve Heisenberg scaling for such a state contain correlators among all the $n$ particles. Concerning the two-anticoherent states instead, as they are close to the state $\ket{\Psi_k}$ for $k= \left\lfloor \frac{n-\sqrt{\frac{n(n+2)}{3}}}{2} \right\rceil$ (even though a bit more difficult to characterize analytically), we would expect a similar conclusion on the impossibility to reach Heisenberg scaling with only the observables in $\vec A_{\mathfrak{su}(2)}$.

\section{Multiparameter $SU(1,1)$ estimation}\label{sec:SU11multi}

\subsection{Single parameter $SU(1,1)$}

Let us consider here the estimation of a phase generated by an $SU(1,1)$ evolution. Even though the $SU(2)$ estimation schemes are more extensively treated in the literature, the $SU(1,1)$ interferometer has been considered in a number of works~\cite{yurke86,Li_2014,Andersonetal2017,Gong_2017}. For better clarity, let us first consider the single-parameter case. First of all, the two-mode bosonic generators obtained from the fundamental representation of $su(1,1)$ are given by:
\be\label{eq:su11algebra}
\begin{aligned}
    K_x&=\frac{1}{2}(a_1^{\dagger} a_2^{\dagger}+a_1 a_2) , \\ 
    K_y&=\frac{1}{2i}(a_1^{\dagger} a_2^{\dagger}-a_1 a_2) , \\ 
    K_z&=\frac{1}{2}(1+a_1^{\dagger}a_1+a_2^{\dagger}a_2) .
\end{aligned}
\ee
We can see that $K_x$ and $K_y$ are two-mode squeezing generators, while $K_z$ is essentially the total number operator.

Thus, a single-parameter $SU(1,1)$ estimation can be the estimation of a real squeezing parameter $\xi$ in an evolution of the form
\be
U(\xi)=e^{-i\xi K_x}.
\ee
In this setting it is clear that the total number of particles will not be conserved, since $SU(1,1)$ operations do not commute with $N$. Thus, the question also arises how to properly identify the resources. An argument similar to the previous section is that the total average number of particles $\ex{N}=:\bar n$ will be still bounded in practical evolutions, since it is related to the total energy of the system. In fact, a typical benchmark for indefinite-$n$ states in the Mach-Zehnder interferometer is provided by a coherent-state input $\probez = \ket{\alpha , 0}$~\cite{parisMetrev09,Demkowicz_Dobrza_ski_2015}, which allows one to reach only shot-noise scaling in terms of the average particle number (see also \cref{sec:indefN}).

Another quantity that might make sense to keep fixed in the context of $SU(1,1)$ estimations, at least from a group-theoretical point of view, is the eigenvalue $m$ of $J_z$, which is a quantity that is left invariant by $SU(1,1)$ unitaries and labels the corresponding irreducible representations. In other words, $SU(1,1)$ unitaries always map eigenstates of $J_z$ with a given eigenvalue $m$ into eigenstates with the same eigenvalue and the question arises as to which sector are states that give the best scaling of precision with $\bar n$.

As a practical example, let us consider the estimation of an infinitesimal squeezing parameter $\xi\simeq 0$ generated by $K_x$ as above. First, it is easy to see that a Fock-state input reaches SQL scaling: 
\be
\begin{gathered}
    \probez= \ket{n,0} \quad \rightarrow \quad 4\var{K_x}_0 = n+1 \\
    \rightarrow \quad \var{\hat \xi}=O(1/n),
\end{gathered}
\ee
while for example once again an input twin-Fock state reaches Heisenberg scaling:
\be
\begin{gathered}
    \probez= \ket{n/2,n/2} \quad \rightarrow \quad 4\var{K_x}_0 = (n^2/2)+n+1 \\ 
    \rightarrow \quad \var{\hat \xi}=O(1/n^2) .
\end{gathered}
\ee
Similarly, it is also easy to see that the NOON state gives a SQL scaling with $\bar n=n$.

Note also how the Fock state has $m=n$, while the twin-Fock state is in the sector $m =0$. Other natural probe states with fluctuating number of particles are again the Heisenberg-Weyl coherent states, as well as squeezed states. The former gives SQL scaling completely analogously as the Fock state because it satisfies
\be
4\var{K_x}_0 = |\alpha|^2+1=\bar{n}+1  ,
\ee
while inputting some initial squeezing allows to reach Heisenberg scaling in $\bar n$. For example, a natural probe state to consider in our context is given by a two-mode squeezed-vacuum
\be
\ket{\Psi_0}=  e^{-i\zeta K_y}\ket{0,0} =\sqrt{1-|q|^2}\sum_n q^n\ket{n,n} \ ,
\ee
where $q=-\tanh(\zeta/2)$. This probe state is also in the sector with $m=0$ and satisfies
\be
4\var{K_x}_0=\bar{n}^2+2\bar{n}+1 \rightarrow \var{\hat \xi}=O(1/\bar{n}^2) ,
\ee
where the average particle number is given by
\be
\bar n =\ex{N}_0 = 2\sinh^2{\frac\zeta2} .
\ee

\begin{figure}[h]
\centering
\includegraphics[width=0.97\columnwidth]{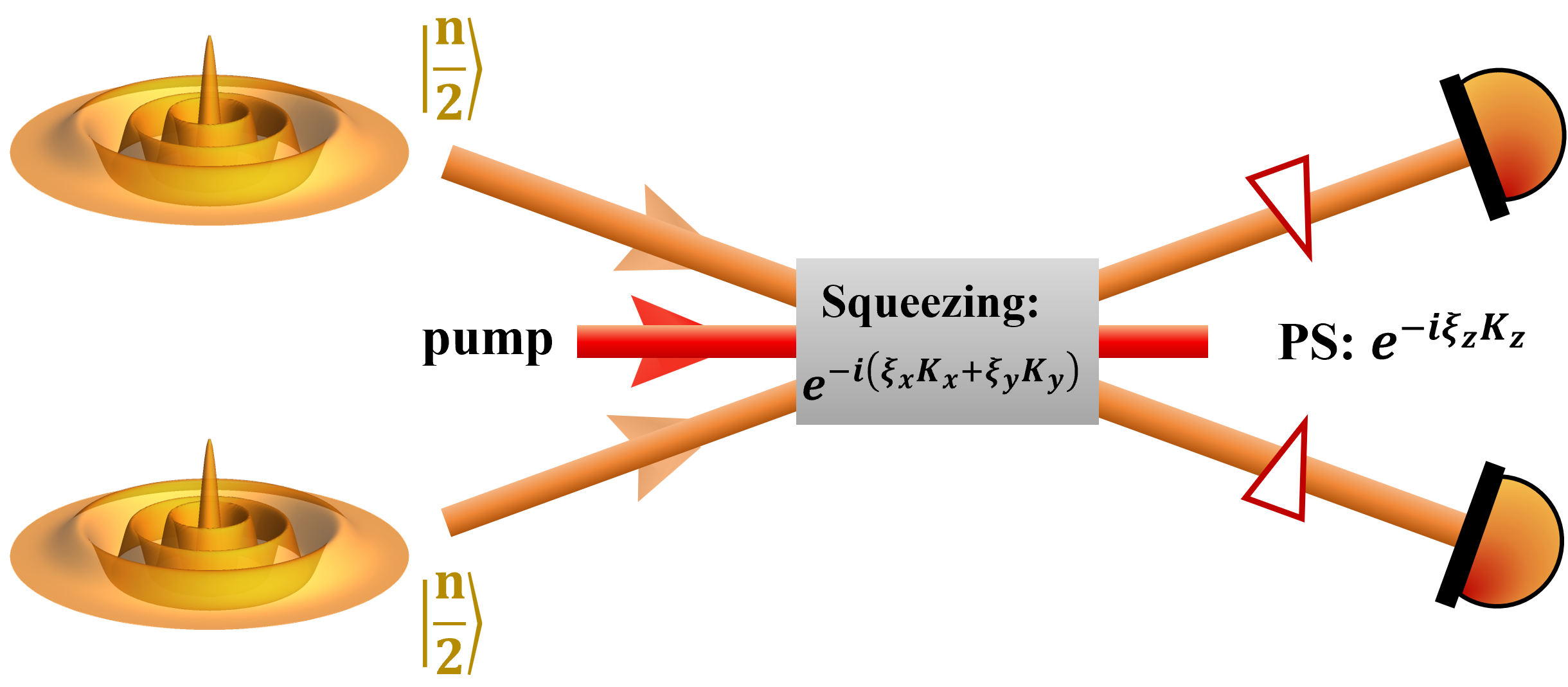} 
\caption{Illustration of a potential interferometric setup designed for three-parameter $SU(1,1)$ estimation. An input two-mode state of the form \cref{eq:su11states} evolves into $U({\vec \xi})\probez$. The pump-driven nonlinear element implements a two-mode $SU(1,1)$ squeezing operation $e^{-i(\xi_x K_x+\xi_y K_y)}$, while a common-mode phase shifter PS realizes $e^{-i\xi_z K_z}$. In the infinitesimal regime this yields \cref{eq:SU11KxKyKzInfinitesimal}. The input is again a twin Fock state, obtained from \cref{eq:su11states} when $k = 0$.}
\label{figPara3SU11}
\end{figure}

Let us now consider the three-parameter estimation of an arbitrary $SU(1,1)$ unitary (cf.~\cref{figPara3SU11}), and let us focus once again on the case of an infinitesimal unitary:
\be\label{eq:SU11KxKyKzInfinitesimal}
U(\vec \xi) \simeq \id - i( \xi_x K_x  + \xi_y K_y + \xi_z K_z) = \id - i \vec \xi \cdot \vec K ,
\ee
which also means that $\tilde{\vec H} = \vec K$ in this case. Here the case of an infinitesimal unitary, besides being possibly the most relevant from a practical perspective, is also the most tractable from a mathematical point of view.

Note once again that there are many different possible parametrizations of such a unitary, and in this case the group of unitaries is not compact, thus preventing an easy treatment of infinitesimal unitaries close to an arbitrary point. See for example \cite{ban} for more details about this group. Nevertheless, we proceed via analogy with the three-parameter $SU(2)$ estimation, trying to emphasize some common traits of the two problems. First, we note that, since $K_z$ is a function of the total number operator, its corresponding phase $\xi_z$ cannot be estimated with probe states that have a definite number of particles, i.e., eigenstates of $N$. Fock states $\ket{n_1,n_2}$ are eigenstates of $K_z$ with eigenvalue $\tfrac 1 2 (1+n_1+n_2)$ and have 
\be\label{eq:QFIMboundTFstatesu11}
\var{K_x}=\var{K_y}=\tfrac 1 4 \left( 1 + n_1 + n_2 +2n_1 n_2 \right) .
\ee
Furthermore, for Fock states the covariance matrix of $K_x,K_y,K_z$ is diagonal. It is clear that if we fix $n_1+ n_2 \leq n$ the best Fock state for estimating the squeezing coefficients $\xi_x\simeq 0$ and $\xi_y\simeq 0$ is given once again by the twin Fock state, which reaches Heisenberg scaling for both phases. 

\subsection{Probe states reaching Heisenberg scaling for three parameters}

Let us now consider superpositions of such Fock states, in the special case $n_1=n_2$, so as to restrict to the subspace $m=0$ which, as we observed, might contain the most useful states. By analogy to the states considered in \cref{eq:probeXYZ} for the $SU(2)$ case, let us consider a state of the form
\be\label{eq:su11states}
 \ket{\tilde \Psi_{k}}:=\frac1{\sqrt2}\left(\ket{\tfrac n 2 , \tfrac n 2} + \ket{\tfrac n 2 +k, \tfrac n 2 +k} \right) ,
\ee
that have an average number of particles given by 
\be
\ex{N}_{\tilde \Psi_{k}} = n+k = \bar n .
\ee
The variances of $K_x$ and $K_y$ are given by
\be
\begin{aligned}
    \var{K_x}_{\tilde \Psi_{k}} &=  \tfrac{1}{8} \left(2 k^2+2 k (n+1)+n (n+2)+2\right) \\
    &=\var{K_y}_{\tilde \Psi_{k}} .
\end{aligned}
\ee

Thus, states of the form in \cref{eq:su11states} can reach Heisenberg scaling for the estimation of $\xi_x$ and $\xi_y$. In order for such a state to reach Heisenberg scaling also for estimating $\xi_z$ we have to look for the one that has $\bar n^2$ spread in the particle number, or in other words, for the state that maximizes $\var{N}$. This will have the form~\footnote{See also \cite{probesNindefinite,probesNindefinite2} for a discussion about the preparation of these and similar states.}
\be
\probez = \ket{\tilde \Psi_{-n/2}} =\tfrac 1 {\sqrt{2}} \left(\ket{0,0} + \ket{n/2,n/2} \right) ,
\ee
which has $\bar n = n/2$ and is such that 
\be
\begin{aligned}
    \var{K_z}_0 &= \frac14 \var{N}_0 =\frac{1}{16} n^2 = \frac 14\bar n^2 , \\
    \var{K_x}_0 &=  \tfrac 1 4 (1+\bar{n}+\bar{n}^2) =\var{K_y}_0 .
\end{aligned}
\ee

Furthermore, once again the covariance matrix of $K_x, K_y$ and $K_z$ is diagonal, and its diagonal entries scale as $O\left(\bar{n}^2\right)$ for all three parameters. Thus, \cref{eq:suffcondsatMCRbound} is met and this state allows in principle for Heisenberg scaling for the full $SU(1,1)$ estimation. Clearly, other classes of states can be considered in the fluctuating-$n$ scenario. See \cref{sec:indefN} for further details and examples. However, as we mentioned here we focus on states of this form which are analogous to those in \cref{eq:probeXYZ}, which we analyzed for the $SU(2)$ estimations. It is also worth noting once more that they are eigenstates of $J_z$ (and $J_z^2$) with eigenvalue $m=0$, but not of $N$.

\subsection{Estimation via multiparameter method of moments}

Following the same approach as in the $SU(2)$ case, we now consider the scenario of an estimation via multiparameter method of moments with the following vector of observables:
\be\label{eq:vecAsu11}
\vec A_{\mathfrak{su}(1,1)} = (\vec K , \tilde{\vec Q} ) ,
\ee
where the vector of generators is given by 
\be
\tilde{\vec H} = \vec K =(K_x,K_y,K_z) ,
\ee
and corresponds to the first three elements, while 
\be
\tilde{\vec Q} = (K_x^2,K_y^2,K_z^2, \{K_x , K_y\} , \{K_x , K_z\} , \{ K_y , K_z \})
\ee
is the vector of quadratic products.

Analogously as in the $SU(2)$ estimation case, for states of the form \eqref{eq:su11states} we have that the expectation values of $K_z$ and $K_z^2$ are insensitive to infinitesimal $SU(1,1)$ unitaries, and therefore we can restrict to the smaller vector
\begin{equation}
    \vec A^\prime_{\mathfrak{su}(1,1)}=(K_x,K_y,\{K_x,K_z\},\{K_y,K_z\})^T ,
\end{equation} 
and again for the same reason we will be able to estimate only phases generated by $K_x,K_y$. We find once again that the twin-Fock state, i.e., the state obtained in $k=0$ from the class in \cref{eq:su11states}, achieves Heisenberg scaling for both phases, and concretely that the accuracy is given by
\be 
\lim_{\boldsymbol{\xi} \rightarrow (0,0,0)}  \COV^{-1}(\hat {\boldsymbol{\xi}}) \leq \nu \, {\rm diag}(\kappa,\kappa,0),
\ee 
where
\be
\kappa=1+\bar n+\tfrac12 {\bar n}^2 ,
\ee
which would in principle allow the QFIM scaling given by \cref{eq:QFIMboundTFstatesu11}. The observables for which one has to estimate the expectation values are given by
\be 
\vec M =(h K_y-\{K_y,K_z\},h K_x-\{K_x,K_z\})^T ,
\ee
where $h=1+\bar n$. Clearly, this is a more abstract scenario, in which the measurements are not necessarily possible to implement in practice. Furthermore, as before it is unclear whether the precision bound that we found can be saturated, even just theoretically, with an imperfect joint measurement of the observables in $\vec M$, since they do not commute (besides being potentially unrealistic to implement). Nevertheless, we do find some analogy between the $SU(1,1)$ and the $SU(2)$ case also in the estimation via second moments of the generators.

\section{Discussion on states with indefinite number of particles}\label{sec:indefN}

\subsection{Indefinite-$n$ states for $SU(2)$ estimation}

Let us now expand briefly the discussion of the case of probe states with indefinite $n$. Typically, accuracy scalings in this case are expressed with respect to $\bar n:= \ex{N}$ instead of $n$. In particular, the shot-noise limit is a scaling with $\bar n$ while the Heisenberg scaling is taken to $\bar n^2$ by analogy to the fixed-$n$ case, and also for practical arguments that rule out faster scalings under realistic circumstances (see, e.g., ~\cite{Demkowicz_Dobrza_ski_2015}). For example, in general to exploit possible particle-number coherences, the term $e^{i\Phi N}$ in the evolution imposes the need of a reference beam~\cite{BartlettRMP2007,JarzynaDobrzankski2012,Demkowicz_Dobrza_ski_2015}. However, formally speaking the Heisenberg scaling is not anymore a strict bound~\cite{Anisimovetal2010}. Note also that, for calculations involving moments of $su(2)$ operators a state of the form
\be
\ket{\Psi_0} = \sum_n c_n \ket{\Psi_n},  
\ee
that is a superposition between states with different $n$ is essentially equivalent to a state 
\be
\varrho_0 = \sum_n |c_n|^2 \ketbra{\Psi_n} ,
\ee
that is just a mixture of states with different $n$. This is because we have
\be
\sum_{n,n^\prime} c_n^* c_{n^\prime} \bra{\Psi_n} J_k^l \ket{\Psi_{n^\prime}} = \sum_n |c_n|^2 \bra{\Psi_n} J_k^l \ket{\Psi_n} ,
\ee
which can also be understood in terms of a superselection rule~\cite{BartlettRMP2007}. In this sense, coherences between different particle numbers are not so crucial for estimations of $SU(2)$ unitaries via moments of $su(2)$ generators~\cite{TothApellaniz2014}. 

Still, one can consider the more general framework of fluctuating-$n$ states, perhaps with physical restrictions on the allowed superpositions and measurements. A typical benchmark for indefinite-$n$ states is provided by a coherent-state input to the Mach-Zehnder interferometer, which allows to reach only shot-noise scaling in terms of the average particle number, i.e., 
\be
\begin{aligned}
    \ket{\Psi_0} &= \ket{\alpha , 0} \quad \rightarrow \\
    \var{J_y}_{\theta\simeq 0} &= \var{J_y}_0 = \tfrac14\bar n \\
    \rightarrow \var{\hat \theta} &\propto \frac 1 {\bar n} ,
\end{aligned}
\ee
where the coherent state is expanded in Fock basis as
\be
\ket{\alpha , 0} = e^{-|\alpha|^2/2}\sum_n\frac{\alpha^n}{\sqrt{n!}}\ket{n,0} ,
\ee
and has an average number of particles given by $\bar n=|\alpha|^2$. 
Thus, the reference for ``classical'' probe states for fluctuating $n$ can be taken as the $SU(2)$ orbit of $\ket{\alpha , 0}$.

Several ideas have been explored to surpass this limit in the fluctuating-$n$ case, for example by making use of squeezed states in one of the input ports~\cite{caves_1981,BondurantShapiro84,yurke86,BrifMann1996,Brif_1996,brif1996high,Gerry2000,pezzesmerziPRL2008,pezzesmerzi2013,Chao-Ping_2016,LiGardetal2016,Youetal2019,Zhangetal2021,DuKongetal2022}.
In this sense, one can see an $SU(1,1)$ operation as a resource for improving the performance of $SU(2)$ interferometers.
In general, it is hard to find the optimal state in the fluctuating-$n$ case, and it is even difficult to make precise statements on the maximum achievable precision scalings, essentially because one can always have a mixture of states with arbitrarily large $n$ but with fixed $\ex{N}$. Similarly, also superpositions 
between states with extremely different particle number give Heisenberg scaling with $\bar n$. As a simple example one can see that 
the following probe state 
\be\label{eq:prstate0n2}
\probez=\left(\frac{\ket{0}+\ket{n}}{\sqrt{2}}\right)\otimes\left(\frac{\ket{0}+\ket{n}}{\sqrt{2}}\right) ,
\ee
has a diagonal QFIM given by
\be
\Qfish{\Psi_\Bth}{} =\tfrac12 {\rm diag}\left(n(n+2),n(n+2),n^2\right) .
\ee
Such a state has $\ex{N}=n$ and thus it allows for Heisenberg scaling in terms of average total particle number for a full three-parameter $SU(2)$ unitary.
A similar conclusion can be made for a three-parameter $SU(1,1)$ unitary, for example considering the state
\be
\probez=\tfrac 1 {\sqrt{2}} \left(\ket{0,0} + \ket{n,n} \right) ,
\ee
as we mentioned earlier.

\subsection{Analysis of common states with indefinite particle number}

\begin{center}
\begin{table*}[t]
\centering
\resizebox{\textwidth}{!}{
\begin{tabular}{|c|c|c|c|c|}
\hline
$\probez$ & $\vec A$ & $\tilde{\vec H}$ & $\mathcal M(\tilde{\vec H})$  & $\vec M$ \\ \hline

$e^{ir L_y}\ket{0,0}$ 
& $(L_y,\vec J)$ & $\vec J$ & ${\rm diag}\left( \bar n , \bar n , \bar n (1+ \bar n)  \right)
$ & $(J_x,J_y,L_y)$ \\ 

\hline

 $e^{ir L_y}\ket{0,0}$ 
 & $(L_y,\vec K)$ & $\vec K$ & ${\rm diag}\left(
    (1+ \bar n) , (1+ \bar n) , 2\bar n (1+ \bar n)
\right)$ & $(K_y,K_x,L_y)$ \\ 

\hline

$e^{ir K_y}\ket{0,0}$ 
& $(L_x,L_y,J_x,J_y)$ & $(J_x,J_y)$ & ${\rm diag}\left(
    \bar n(2+ \bar n) , \bar n (2+ \bar n)\right)$ & $\left(J_x+2\sqrt{\frac{|\bar n-1|}{1+ \bar n}} L_x, J_y+2\sqrt{\frac{|\bar n-1|}{1+ \bar n}} L_y\right)$ \\ 
    
    \hline
    
$\ket{\alpha,0}$ 
& $(L_x,L_y,\vec J)$ & $\vec J$ & 
${\rm diag}\left(
    \bar n, \bar n , \frac{2 \bar n^2}{2\bar n+1} \right)$
& $\left(J_x,J_y,\frac{\Im(\alpha^2)}{\Re(\alpha^2)}L_x+L_y\right)$ \\ 

\hline

$\ket{\alpha,0}$ 
& $(L_x,L_y,\vec K)$ & $\vec K$ & 
${\rm diag}\left(
    \bar n+1, \bar n +1 , \frac{2 \bar n^2}{2\bar n+1} \right)$
 & $\left(K_x,K_y,\frac{\Im(\alpha^2)}{\Re(\alpha^2)}L_x+L_y\right)$ \\ 
 
 \hline

\end{tabular}
}
\caption{Summary of multiparameter $SU(2)$ and $SU(1,1)$ estimations with typical Gaussian states as inputs}\label{table1}
\end{table*}  
\end{center}

A paradigmatic class of probe states with indefinite particle number is given by Gaussian states. In fact, both the $SU(2)$ and the $SU(1,1)$ unitaries belong to the class of Gaussian unitaries, i.e., those that have a Hamiltonian quadratic in the mode operators. These unitaries always transform Gaussian states into other Gaussian states and have in fact often been investigated in the context of quantum metrology with Gaussian resources~\cite{PinalPRA2013,FriisPRA2015,Sparaciari_15,SparaciariOlivaresParis,SafranekFuentes2016,Rigovaccaetal2017,Nichols_2018,Bakmou_2020,sorelli2023gaussian}. In fact, it is a natural question to ask what is the precision limit that can be reached with only Gaussian probe states and encoding evolutions, plus measurement observables quadratic in the mode operators. In our framework this can be also seen as a way to look at the interplay between $SU(2)$ and $SU(1,1)$ operations, which would generalize the idea of improving $SU(2)$ estimations with a prior $SU(1,1)$ operation, pushed forward by seminal works~\cite{caves_1981,BondurantShapiro84,yurke86}, and later extended in a broad framework~\cite{BrifMann1996,Brif_1996,brif1996high,Gerry2000,Chao-Ping_2016,LiGardetal2016,Youetal2019,Zhangetal2021,DuKongetal2022}. Besides two-mode operators forming $su(2)$ or $su(1,1)$ basis, we consider also single-mode $su(1,1)$ operators, that have the form
\be
\begin{aligned}
    L_x &=\frac 1 4 (a^\dagger a^\dagger+aa) , \\
    L_y &=\frac 1 {4i} (a^\dagger a^\dagger-aa), \\
    L_z &=\frac1 4 (a^\dagger a+aa^\dagger) ,
\end{aligned}
\ee
and similar definitions for mode $b$. Note that essentially the operator $L_z$ corresponds to the number of particles in mode $a$. As observed often in literature in different contexts~\cite{caves_1981,BondurantShapiro84,yurke86,pezzesmerziPRL2008,GAIBA2009934,pezzesmerzi2013}, unitaries generated by these operators (i.e., single-mode squeezing operations) are also important resources for parameter estimations. However, it is not essential to consider single-mode squeezing operations in both modes, and thus for simplicity we only consider one set
of such single mode $su(1,1)$ operators.
Thus, overall we consider the following vector of observables 
\be
\vec A_{\rm G}= (L_x,L_y,L_z,K_x,K_y,K_z,J_x,J_y,J_z ) 
\ee
and various probe Gaussian states. 
As we did previously, whenever an observable does not contribute to the estimation or leads to a singular moment matrix, we drop it from the set.
In \cref{table1} we summarize our findings with concrete probe states and measurements. 

As an overall summary, we conclude that single- and two-mode squeezed states represent good probe states for both $SU(2)$ and $SU(1,1)$ estimations, which is something in agreement with well-established results~\cite{caves_1981,BondurantShapiro84,yurke86,pezzesmerziPRL2008,GAIBA2009934,Anisimovetal2010,pezzesmerzi2013,Li_2014,Gong_2017,fadel2024quantummetrologycontinuousvariable}. We also observe that it is somehow crucial to allow measurements of the observables $L_x,L_y$. As a final comment, and as a term of comparison in this context, we also analyze other highly quantum states with fluctuating-$n$ that are often considered in literature, such as Schr\"odinger-cat-like states~\cite{JooMunroSpiller2011,Liu_2016,Chao-Ping_2016,fadel2024quantummetrologycontinuousvariable}. It might be perhaps counter-intuitive that we find (for various forms of cat-like states) that only shot-noise scaling can be achieved in our context, i.e., by using the method of moments with the vector of observables $\vec A_G$. However, intuitively, our findings just confirm that in this case more complicated observables would have to be included, as it is the case for NOON states or analogous ones.

\section{Conclusion and outlook}\label{sec:conclusions}

In conclusion, we have investigated the multiparameter estimation of $SU(2)$ and $SU(1,1)$ unitaries and the limit to their precision scaling. First, for each of the two estimation tasks we identified a class of pure states containing ideal probes that would potentially allow for reaching Heisenberg precision scaling simultaneously in all three parameters according to the quantum Fisher information matrix and the relative multiparameter quantum Cram\'er-Rao bound. In the $SU(2)$ case, we consider subspaces (sectors) of states with fixed particle number, as the operator $N$ itself is left invariant by the parameter-encoding evolution. Furthermore, the value of $\ex{N}=n$ can be seen as a resource connected with the total energy of the system. As an ideal class of probe states we identified eigenstates of $J^2_z$ with different fixed-$n$ eigenvalues. Such a class contains paradigmatic states, such as the NOON state, the twin-Fock state and also states close to those which have been called two-anti-coherent in recent literature~\cite{Bouchard:17,ChryssomalakosCoronado2017,GoldbergJames2018,Martin2020optimaldetectionof,Z_Goldberg_2021,Serrano_Ens_stiga_2025}.  

In the $SU(1,1)$ case instead, the particle number is not anymore fixed by the evolution, but the analogous invariant becomes the value of $J_z$. Still, the average particle number, even if not conserved by the encoding evolution, can be seen as the total energy resource of the system. Therefore, we consider subspaces of states with different values of $\ex{J_z}=m$ and identify a class of states in the $m=0$ sector that contains states that would allow Heisenberg scaling of precision in terms of $\ex{N} = \bar n$ in all three parameters according to their QFIM. Again, motivated by the fact that the sector with $m=0$ contains ideally useful states, we consider a class of eigenstates of $J_z^2$ which is analogous to the $SU(2)$ case.

Afterwards, we considered a more pragmatic scenario in which the estimation is performed via signal-to-noise ratio of first and second moments of the ($su(2)$ and $su(1,1)$) generators on the probe states. Analyzing such a scenario with the ideal class of pure probe states that we have identified, we found that the twin-Fock state emerges in both $SU(2)$ and $SU(1,1)$ estimations as the only state that allows one to achieve Heisenberg limit for at least two out of the three parameters. As a comparison, we also investigated the case of other paradigmatic probe states with a fluctuating particle number, such as Gaussian states and cat states, allowing only the measurement of expectation values of operators quadratic in the mode operators. In this case we find that only the two-mode squeezed state emerges for allowing simultaneous Heisenberg scaling in a two-parameter $SU(2)$ estimation, which confirms the resourceful nature of such a state as discussed in the single-parameter scenario~\cite{caves_1981,BondurantShapiro84,yurke86} and also generalizes to the multiparameter case the arguments on the enhancement of $SU(2)$ estimations with prior $SU(1,1)$ operations~\cite{BrifMann1996,Brif_1996,brif1996high,Gerry2000,Chao-Ping_2016,LiGardetal2016,Youetal2019,Zhangetal2021,DuKongetal2022}.  

As an outlook, we anticipate that our methods are readily generalizable to unitary groups containing more parameters, such as the $SU(d)$ and the $SU(p,q)$ groups. Also in those cases it is possible to identify ideal classes of states in sectors left invariant by the encoding evolution and analyze the estimation precision with the error-propagation formula from measurements of (low) powers of the generators. Still, the open question remains on how to saturate the asymptotic bounds coming from the method of moments when the measured observables do not commute and one is forced to do either independent or imperfect measurements of the noncommuting observables. We believe these and related questions, for example, further connections between optimal probe states for larger subgroups of unitaries, and the notion of anti-coherent states~\cite{Serrano_Ens_stiga_2025}, would deserve further investigation. Moreover, from a practical point of view, it could be interesting to explore a more general framework, for example including non-ideal preparations, practical measurement schemes, as well as going towards global estimations (at least in the $SU(2)$ or other compact groups cases) and consider a Bayesian approach. We thus hope that our work could trigger subsequent investigation with group-theoretic methods in the estimation of multiparameter unitaries.

{\bf Acknowledgements.---} This work is supported by the National Natural Science Foundation of China (Grant No. 12405005). SL acknowledges the China Postdoctoral Science Foundation (No. 2023M740119). FESS acknowledges the institutional projects $376/2020$ and $475/2023$ from Federal University of Mato Grosso, partial funding from the brazilian agency FINATEC and the grant PTRES 170136, FONTE 1050A000AP, UGR 154228, PI MGY01N0104N from University of Bras\'ilia. GV acknowledges financial support from the Austrian Science Fund (FWF) through the grants P 35810-N and P 36633-N (Stand-Alone) and from the grant RYC2024-048278-I funded by MCIU/AEI/10.13039/501100011033 and FSE+.

\appendix

\begin{widetext}

\section{Details on the method-of-moments estimator}\label{app:MoMdetails}

In this appendix we briefly recap the method of moments (MoM) in a form suited to our setting.

\subsection{Single-parameter case}

Let $M= M(q)$ be a random variable (e.g., the value of an observable in a projective measurement), and let each single-shot outcome $q$ be distributed as ${\rm p}(q|\theta)$. Define the sample mean after $\nu$ shots:
\begin{equation}
\bar M_\nu:=\frac{1}{\nu}\sum_{t=1}^\nu m_{q_t} .
\end{equation}
Assume that the theoretical mean
\begin{equation}
\mu(\theta):=\mathbb{E}_\theta[M]=\sum_q m_q\, {\rm p}(q|\theta)
\end{equation}
is known and (locally) invertible. The MoM estimator is defined implicitly by matching the empirical and theoretical moments,
\begin{equation}
\mu(\hat\theta)=\bar M_\nu
\qquad\Longrightarrow\qquad
\hat\theta=\mu^{-1}(\bar M_\nu).
\end{equation}
For large $\nu$, the central limit theorem (CLT) implies that the random variable $\sqrt{\nu}\,(\bar M_\nu-\mu(\theta))$ is normally distributed with zero mean and variance given by
\begin{equation}
\mathrm{Var}_\theta [\bar M_\nu]=\mathbb{E}_\theta[M^2]-\mu^2(\theta) .
\end{equation}
Then, from a first-order Taylor expansion one obtains
\begin{equation}
\hat\theta-\theta \;\simeq\; \frac{1}{\mu'(\theta)}\bigl(\bar M_\nu -\mu(\theta)\bigr),
\end{equation}
where $\mu'(\theta):=\partial_\theta \mu(\theta)$. 

Because of this, the estimator is asymptotically unbiased and has a mean-squared error (MSE) given by
\begin{equation}
\mathrm{MSE}[\hat\theta]
=\mathbb{E}_\theta[(\hat\theta-\theta)^2]
\;=\;\frac{1}{\nu}\,\frac{\mathrm{Var}_\theta [M]}{\mu'(\theta)^2}+O(\nu^{-2}).
\label{eq:mom_single_mse}
\end{equation}
In the quantum setting, if $M$ corresponds to a Hermitian operator measured on a density matrix $\varrho_\theta$, then the mean is $\mu(\theta)=\langle M\rangle_{\varrho_\theta}$ and the variance is $\mathrm{Var}_\theta[M] = (\Delta M)^2_{\varrho_\theta}$.

Unbiased estimators satisfy the (classical) Cram\'er--Rao bound,
\begin{equation}
\mathrm{MSE}[\hat\theta]\ge \frac{1}{\nu\,\mathcal{F}(\theta)},
\end{equation}
where $\mathcal{F}(\theta)$ is the Fisher information for a single shot. Pre-asymptotic corrections to \eqref{eq:mom_single_mse} typically enter at $O(\nu^{-2})$ and depend on higher cumulants as well as higher derivatives of $\mu(\theta)$; see, e.g., Refs.~\cite{kolassa2006series,HervasetalPRL2025}. The method of moments estimator is asymptotically unbiased~\cite{pezze2014quantum} so that in this regime the MSE coincides with the variance $(\Delta \thetaMoM)^2$.

\subsection{Multiparameter case}\label{app:multiMoM}

Let $\vec\theta=(\theta_1,\dots,\theta_r)$ be the unknown parameter vector. First, consider a \emph{single joint measurement} described by a PVM $\{\ketbra{q}\}$, and let $\vec M=(M_1,\dots,M_k)$ be a collection of \emph{commuting} Hermitian operators that are diagonal in that same basis,
\begin{equation}
M_i=\sum_q m_i(q)\ketbra{q},
\qquad [M_i,M_j]=0.
\end{equation}
The joint outcome $q$ is distributed as ${\rm p}(q|\vec\theta)=\mathrm{Tr}[\varrho_{\vec\theta}\ketbra{q}]$, and it induces a classical random vector
\begin{equation}
\vec M(q):=\bigl(m_1(q),\dots,m_k(q)\bigr)^T.
\end{equation}
The empirical mean vector and the theoretical mean vector are
\begin{equation}
\bar{\vec M}_\nu:=\frac{1}{\nu}\sum_{t=1}^\nu \vec M(q_t),
\qquad
\vec\mu(\vec\theta):=\mathbb{E}_\theta[\vec M]
=\bigl(\langle M_1\rangle_{\varrho_{\vec\theta}},\dots,\langle M_k\rangle_{\varrho_{\vec\theta}}\bigr)^T.
\end{equation}
The MoM estimator is defined by the vector equation
\begin{equation}
\vec\mu(\hat{\vec\theta})=\bar{\vec M}_\nu .
\label{eq:mom_multi_def}
\end{equation}

Define the $k\times r$ Jacobian
\begin{equation}
C_{ij}(\vec\theta):=\partial_{\theta_j}\mu_i(\vec\theta)=\partial_{\theta_j}\langle M_i\rangle_{\varrho_{\vec\theta}}.
\label{eq:C_def}
\end{equation}
The multivariate CLT yields that the vector of random variables $\sqrt{\nu}\,(\bar{\vec M}_\nu -\vec\mu(\vec\theta))$ is normally distributed with mean zero and covariance matrix given by $\mathrm{Cov}_{\vec \theta}(\vec M)$. Linearizing \eqref{eq:mom_multi_def} around $\vec\theta$ gives
\begin{equation}
\hat{\vec\theta}-\vec\theta \;\simeq\; C^{+}(\vec\theta)\bigl(\bar{\vec M}-\vec\mu(\vec\theta)\bigr),
\end{equation}
where $C^{+}$ denotes a left-inverse (for $k=r$ and $\det C\neq 0$, this is $C^{-1}$). Hence, once again the MoM is asymptotically unbiased. Its MSE matrix following from the CLT (i.e., valid in the large $\nu$ limit) is given by:
\begin{equation}
\mathrm{MSE}[\hat{\vec\theta}]
:=\mathbb{E}_\theta\bigl[(\hat{\vec\theta}-\vec\theta)(\hat{\vec\theta}-\vec\theta)^T\bigr]
=\frac{1}{\nu}\,C^{+}\,\mathrm{Cov}_{\vec \theta}[\vec M] \,\bigl(C^{+}\bigr)^{T}+O(\nu^{-2}).
\label{eq:mom_multi_mse}
\end{equation}

In this case of commuting observables, since each repetition produces a joint outcome $q_t$ that gives the value of all observables, the empirical covariance of the sampled statistics converges to the covariance matrix of the measured observables, i.e., we have:
\begin{align}
[\mathrm{Cov}_{\vec \theta}[\vec M]]_{ij}
&:=\frac{1}{\nu-1}\sum_{t=1}^\nu \bigl(m_i(q_t)- (\bar M_\nu)_i\bigr)\bigl(m_j(q_t)-(\bar M_\nu)_j\bigr) 
=\langle M_i M_j\rangle_{\varrho_{\vec\theta}}-\langle M_i\rangle_{\varrho_{\vec\theta}}\langle M_j\rangle_{\varrho_{\vec\theta}} = [\Gamma_{\varrho_{\vec \theta}}[\vec M]]_{ij}.
\label{eq:Gamma_def}
\end{align}
If $\varrho_{\vec\theta}=U(\vec\theta)\varrho\,U^\dagger(\vec\theta)$ and we work in the Heisenberg picture $M_i(\vec\theta)=U^\dagger(\vec\theta)M_iU(\vec\theta)$, then
\begin{equation}
C_{ij}(\vec\theta)=\partial_{\theta_j}\langle M_i(\vec\theta)\rangle_{\varrho}
=-\mathrm{i}\,\bigl\langle [M_i(\vec\theta),\tilde H_j(\vec\theta)]\bigr\rangle_{\varrho} := [\Omega_\varrho[\vec M(\vec \theta) , \tilde{\vec H}(\vec \theta)] ]_{ij},
\end{equation}
with $\tilde H_j(\vec\theta):=\mathrm{i}\,U^\dagger(\vec\theta)\,\partial_{\theta_j}U(\vec\theta)$, which connects the Jacobian directly to commutators with the (local) generators $\tilde H_j(\vec\theta)$. Here we also defined the commutator matrix $\Omega_{\varrho_{\vec \theta}}[\vec M , \tilde{\vec H}]=\Omega_\varrho[\vec M(\vec \theta) , \tilde{\vec H}(\vec \theta)]$. Compactly, one can introduce the moment matrix
\begin{equation}
\mathcal{M}_{\varrho_{\vec\theta}}[\vec M]:=\Omega^T_{\varrho_{\vec \theta}}[\vec M , \tilde{\vec H}]\,\Gamma^{-1}_{\varrho_{\vec \theta}}[\vec M] \, \Omega_{\varrho_{\vec \theta}}[\vec M , \tilde{\vec H}],
\label{eq:moment_matrix_def}
\end{equation}
such that the MSE matrix can be written as~\cite{gessnerNatComm20}
\be\label{eq:MSEfromMOM}
\mathrm{MSE}[\hat{\vec\theta}]
=\frac{1}{\nu}\,\mathcal{M}^{-1}_{\varrho_{\vec\theta}}[\vec M]+O(\nu^{-2}).
\ee

Note that this argument can be extended also to the case of {\it non-commuting} quantum observables $\vec M$. With the same reasoning, one can resort to the CLT to find the asymptotic expression for the MSE matrix as in \cref{eq:MSEfromMOM}. In that case, we have to consider the {\it real} covariance matrix, which is given by the symmetrized version $[\Gamma_{\varrho_{\vec \theta}}[\vec M]]_{ij} = \tfrac 1 2 \langle M_i M_j + M_j M_i\rangle_{\varrho_{\vec\theta}}-\langle M_i\rangle_{\varrho_{\vec\theta}}\langle M_j\rangle_{\varrho_{\vec\theta}}$. However, when the observables of interest do \emph{not} commute, one cannot in general associate them to a single joint PVM on the original system. A natural alternative is to perform an \emph{imperfect joint measurement} described by a (generally non-projective) POVM; operationally, this produces joint classical data from which one may extract (noisy) estimates. An exemplary scenario is the extraction of Husimi type of quasi-probability distributions, as it is done via heterodyne measurements of quadrature operators, or non-orthogonal projectors onto spin-coherent states for the case of spin variables.

Let $\{E_u\}$ be such a noisy POVM, where $u$ can be discrete or continuous, and we use a different letter to emphasize that it is not associated to the spectrum of the quantum operators $\vec M$ that we want to measure. In this case, the outcome probability is given again by the Born rule:
\begin{equation}
{\rm p}(u|\vec\theta)=\tr[\varrho_{\vec\theta}E_u] ,
\end{equation}
and we can take a vector of (real) processing functions $\vec M(u)=(m_1(u),\dots,m_k(u))^T$ to define the classical random vector
\begin{equation}
\vec M := \vec M(u) ,
\end{equation}
and repeat the same reasoning as above to define the MoM estimator and find its asymptotic unbiasedness and its MSE.

From $\nu$ independent repetitions we obtain outcomes $q_1,\dots,q_\nu$ and hence the sampled mean vector
\begin{equation}
\bar{\vec M}_\nu:=\frac{1}{\nu}\sum_{t=1}^\nu \vec M(q_t) ,
\end{equation}
analogously as before, which defines the MoM estimator. The theoretical mean can be also still computed analogously as before:
\be
\vec\mu(\vec\theta) :=\mathbb{E}_{\vec\theta}[\vec M] =  \int {\rm d} u\,\vec m(u)\, {\rm p}(u|\vec\theta),
\ee
where we used a continuous variable notation for the outcomes for generality.

Similar relation holds for other moments like
\be
\mean{M_i M_j}_{\varrho_\theta} = \int {\rm d} u\, m_i(u) m_j(u) \, {\rm p}(u|\vec\theta) ,
\ee
which can be expressed also by defining operators associated to the moments in terms of the POVM elements: 
\begin{equation}
F_i := \int {\rm d} u \, m_i(u)\,E_u,
\qquad
F_{ij} := \int {\rm d} u \, m_i(u) \, m_j(u) \,E_u ,
\label{eq:povm_moment_ops}
\end{equation}
so that 
\begin{equation}
\mu_i(\vec\theta)=\tr[\varrho_{\vec\theta}F_i],
\qquad
[\Gamma_{\varrho_\theta}(\vec M)]_{ij}=\tr[\varrho_{\vec\theta}F_{ij}]-\tr[\varrho_{\vec\theta}F_i]\tr[\varrho_{\vec\theta}F_j].
\end{equation}

If the goal is to jointly infer expectation values of a non-commuting operator vector $\vec A=(A_1,\dots,A_k)$, a standard notion of an \emph{unbiased} (first-moment preserving) joint measurement is
\begin{equation}
F_i = A_i\qquad \text{for all }i.
\label{eq:unbiased_povm}
\end{equation}
In this case, the POVM outcomes reproduce the target first moments exactly,
\begin{equation}
\mu_i(\vec\theta)=\tr[\varrho_{\vec\theta}A_i]=\langle A_i\rangle_{\varrho_{\vec\theta}}.
\end{equation}
The price for joint access to non-commuting observables is then paid at the level of second moments: the measurement is necessarily noisy, and the covariance of the sampled statistics contains additional noise~\cite{busch2016quantum}.

Alternatively, one may estimate different components by measuring different POVMs on disjoint subsets of the $\nu$ copies. In this case, the corresponding sample means are formed from \emph{independent} datasets, which in turn requires a large number of samples for each independent observable, as well as additional ones to estimate the MSE from the covariances.  

\subsubsection{Example: heterodyne measurement of two quadratures}

A canonical way to \emph{jointly, but imperfectly} access two non-commuting quadratures is heterodyne detection on a single bosonic mode. Let $a$ be the annihilation operator and define dimensionless quadratures
\begin{equation}
x:=\frac{a+a^\dagger}{\sqrt{2}},\qquad
p:=\frac{a-a^\dagger}{i\sqrt{2}},
\qquad [x,p]=i \id.
\end{equation}
The heterodyne POVM is the coherent-state POVM
\begin{equation}
E(\alpha)=\frac{1}{\pi}\ketbra{\alpha},\qquad \alpha\in\mathbb{C},\qquad 
\int_{\mathbb{C}} d^2\alpha\,E(\alpha)=\id ,
\end{equation}
so that the outcome distribution is the Husimi $Q$ function:
\begin{equation}
{\rm p}(\alpha|\vec\theta)=\tr[\varrho_{\vec\theta}E(\alpha)]
=\frac{1}{\pi}\bra{\alpha}\varrho_{\vec\theta}\ket{\alpha}
=: \mathcal Q_{\varrho_{\vec\theta}}(\alpha).
\end{equation}
Writing the complex outcome as
\begin{equation}
\alpha=\frac{x+i p}{\sqrt{2}}
\qquad\Longleftrightarrow\qquad
x=\sqrt{2}\,\Re\alpha,\;\; p=\sqrt{2}\,\Im\alpha,
\end{equation}
each experimental run returns a \emph{pair} $\alpha_t = (x_t,p_t)$ from a single shot. We then define the empirical means and empirical covariance matrix from the data stream
\begin{equation}
\bar x=\frac{1}{\nu}\sum_{t=1}^\nu x_t,\qquad
\bar p=\frac{1}{\nu}\sum_{t=1}^\nu p_t,
\end{equation}
\begin{equation}
\widehat{\Gamma}
=
\begin{pmatrix}
\widehat{\mathrm{Var}}(x) & \widehat{\Cov}(x,p)\\
\widehat{\Cov}(p,x) & \widehat{\mathrm{Var}}(p)
\end{pmatrix},
\qquad
\widehat{\Cov}(x,p)
:=\frac{1}{\nu-1}\sum_{t=1}^\nu (x_t-\bar x)(p_t-\bar p),
\end{equation}
(and similarly for $\widehat{\mathrm{Var}}(x),\widehat{\mathrm{Var}}(p)$). Since each repetition produces \emph{joint} outcomes $(x_t,p_t)$, we can resort to the law of large numbers to assume
\begin{equation}
\bar x \xrightarrow{\nu\to\infty} \mathbb{E}_{\vec\theta}[x],\qquad
\bar p \xrightarrow{\nu\to\infty} \mathbb{E}_{\vec\theta}[p],
\qquad
\widehat{\Gamma}\xrightarrow{\nu\to\infty}\Gamma_{\varrho_{\vec\theta}}(\vec X),
\end{equation}
where $\Gamma_{\varrho_{\vec\theta}}(\vec X)$ can be also drawn from heterodyne statistics, i.e., from the moments:
\begin{equation}
\mathbb{E}_{\vec\theta}[f(x(\alpha),p(\alpha))]=\int_{\mathbb{C}} {\rm d}^2\alpha\, f(x(\alpha),p(\alpha))\, {\rm p}(\alpha|\vec\theta) ,
\end{equation}
for suitable functions $f(x(\alpha),p(\alpha))$. The $Q$ distribution reproduces \emph{first moments} of the quadratures exactly:
\begin{equation}
\mathbb{E}_{\vec\theta}[\vec X]=\langle \vec X\rangle_{\varrho_{\vec\theta}} .
\label{eq:het_first_moments}
\end{equation}
For \emph{second moments}, heterodyne detection is a noisy joint measurement: it adds one half quantum of vacuum noise to each marginal quadrature. Concretely,
\begin{equation}
\mathrm{Var}_{\vec\theta}(x)=\mathrm{Var}_{\varrho_{\vec\theta}}(X)+\frac{1}{2},\qquad
\mathrm{Var}_{\vec\theta}(p)=\mathrm{Var}_{\varrho_{\vec\theta}}(P)+\frac{1}{2}.
\label{eq:het_added_noise}
\end{equation}
Equivalently, one may think of the heterodyne outcomes as an effective classical model
\begin{equation}
x = X + \xi_x,\qquad p = P + \xi_p,
\end{equation}
where $\xi_x,\xi_p$ are independent, zero-mean Gaussian noises with variance $1/2$ (the ``vacuum penalty''), independent of the state.

For the \emph{cross covariance}, no additional constant term appears; rather one obtains the symmetrized quantum covariance:
\begin{equation}
\Cov_{\vec\theta}(x,p)
=\frac{1}{2}\bigl\langle XP+PX\bigr\rangle_{\varrho_{\vec\theta}}
-\langle X\rangle_{\varrho_{\vec\theta}}\langle P\rangle_{\varrho_{\vec\theta}}
\;=: \Cov_{\varrho_{\vec\theta}}^{(\mathrm{sym})}(X,P).
\label{eq:het_cross_cov}
\end{equation}
Thus, the heterodyne covariance matrix can be written compactly as
\begin{equation}
\Gamma(\vec\theta)
=
\Gamma^{(\mathrm{sym})}_{\varrho_{\vec\theta}}(X,P)
+\frac{1}{2}\,\id_2,
\qquad
\Gamma^{(\mathrm{sym})}_{\varrho}(X,P)
:=
\begin{pmatrix}
\mathrm{Var}_\varrho(X) & \Cov^{(\mathrm{sym})}_\varrho(X,P)\\
\Cov^{(\mathrm{sym})}_\varrho(P,X) & \mathrm{Var}_\varrho(P)
\end{pmatrix}.
\label{eq:het_Gamma_decomp}
\end{equation}

\subsection{Optimization over an accessible operator set and attainability of the moment-matrix bound}
\label{app:mom_optimal_bound}

In many practical scenarios one has access only to a restricted set of observables
\begin{equation}
\vec A := (A_1,\dots,A_n)^T,
\end{equation}
from which both the (local) generators and the measured observables are chosen as linear combinations. In particular, we write (local) generators as
\begin{equation}
\tilde{\vec H}(\vec\theta) = R(\vec\theta)\,\vec A,
\label{eq:H_expand_Rtheta}
\end{equation}
with a (generally $\vec\theta$-dependent) real matrix $R(\vec\theta)$, and the measured observables as
\begin{equation}
\vec M = S \vec A,
\label{eq:M_expand_S}
\end{equation}
for some real matrix $S$.

Given a state $\varrho$, define the covariance matrix of $\vec A$ and the commutator matrix
\begin{equation}
\bigl[\Cov_\varrho(\vec A)\bigr]_{kl}:=\Cov_\varrho(A_k,A_l),
\qquad
\bigl[\Omega_\varrho(\vec A)\bigr]_{kl}:=-\mathrm{i}\,\langle[A_k,A_l]\rangle_\varrho,
\end{equation}
(where for noncommuting operators we take the \emph{symmetrized} covariance
$\Cov_\varrho(A_k,A_l):=\tfrac12\langle A_kA_l+A_lA_k\rangle_\varrho-\langle A_k\rangle_\varrho\langle A_l\rangle_\varrho$, which corresponds to the real part of the non-symmetrized covariance matrix).
The associated ``accessible'' moment matrix is
\begin{equation}
\mathcal{M}_\varrho[\vec A]
:=\Omega^T_\varrho(\vec A) \;\Cov^{-1}_\varrho(\vec A)\;\Omega_\varrho(\vec A).
\label{eq:accessible_moment_matrix_app}
\end{equation}

For any choice of measured observables $\vec M$ and generators $\tilde{\vec H}$ of the form
\eqref{eq:H_expand_Rtheta}--\eqref{eq:M_expand_S}, the moment matrix associated with the MoM protocol obeys the bound
\begin{equation}
\mathcal{M}_\varrho\bigl[\tilde{\vec H}(\vec\theta),\vec M\bigr]
\;\le\;
R(\vec\theta)\,\mathcal{M}_\varrho[\vec A]\,R^T(\vec\theta),
\label{eq:moment_matrix_upper_bound_app}
\end{equation}
as shown in Ref.~\cite{gessnerNatComm20}.
Moreover, the bound is saturated whenever there exists an invertible matrix $G$ such that
\begin{equation}
G\,S \;=\; R(\vec\theta)\,\Omega_\varrho(\vec A)\,\Cov^{-1}_\varrho(\vec A),
\label{eq:GS_condition}
\end{equation}
in which case one may choose an optimal set of observables
\begin{equation}
\vec M_{\rm opt}
=
G^{-1}\,R(\vec\theta)\,\Omega_\varrho(\vec A)\,\Cov^{-1}_\varrho(\vec A)\,\vec A.
\label{eq:Mopt_general}
\end{equation}
The key point is that Eq.~\eqref{eq:moment_matrix_upper_bound} is an \emph{algebraic} optimization over linear combinations of the accessible operators and is therefore independent of how (or whether) the components of $\vec M_{\rm opt}$ can be measured jointly in a single shot.

Whether the upper bound \eqref{eq:moment_matrix_upper_bound} is operationally saturable depends on the measurement model, i.e.\ on what is allowed in collecting the classical data used to form the empirical moments.
The bound \eqref{eq:moment_matrix_upper_bound} is automatically attainable by a single PVM whenever the optimal observables commute. If the optimal observables do not commute, then attainability becomes model-dependent: a single-shot joint POVM typically adds noise (hence one may fail to saturate the bound), while separate measurements avoid joint-measurement noise but require a larger number of samples.

Beyond the CLT regime, the covariance admits a systematic expansion in $1/\nu$, whose next-to-leading terms depend on higher-order cumulants of the measurement outcomes and on higher derivatives of the mean-value map $\vec\mu(\vec\theta)=\langle\vec M(\vec\theta)\rangle$. Such pre-asymptotic corrections can be addressed using higher-order asymptotic methods (see, e.g., \cite{kolassa2006series,mccullagh2018tensor,braunstein1992large}) and, in the quantum setting, may be viewed as multiparameter generalizations of corrections beyond the QCRB~\cite{HervasetalPRL2025}. In the present work we restrict to the leading $1/\nu$ behavior captured by the moment-matrix formalism.

\section{Recap of single-parameter $SU(2)$ estimation}\label{app:recapsinglesu2}

Here, we reconsider some known results about the precision limits of estimation schemes where the evolution belongs to $SU(2)$. 
As we mentioned, it is well known that the Mach-Zehnder Interferometer can be described in terms of the Lie-group $SU(2)$~\cite{yurke86,caves_1981,KitagawaUeda1993,Wineland1994Squeezed}. Mathematically, applying the so-called Jordan-Schwinger map
to the spin-$1/2$ operators $J_k=\mathbf{a}^{\dagger}\cdot \tfrac 1 2 \sigma_k\cdot\mathbf{a}$ where $\sigma_k$ are the usual Pauli matrices with $k=x,y,z$ and $\mathbf{a}=(a_1,a_2)^T$, we obtain
\be
\begin{aligned}
    J_x&=\frac{1}{2}(a^{\dagger}_1a_2+a_1a^{\dagger}_2) \\ 
    J_y&=\frac{1}{2i}(a^{\dagger}_1a_2-a_1a^{\dagger}_2) \\  
    J_z&=\frac{1}{2}(a^{\dagger}_1a_1 - a^{\dagger}_2a_2) , \label{spinop_app}
\end{aligned}
\ee
which generate a $su(2)$ algebra.
In an optical setup, $J_x$ and $J_y$ correspond to the interaction of a beam-splitter and in general two-mode systems (e.g., a Bose-Einstein Condensate with two internal states or in a double-well), these operators correspond to tunneling between the modes. In both cases, $J_z$ is half the population difference between modes. 

The $su(2)$ operators are observables that can be measured via the action of a suitable $SU(2)$ unitary followed by measurements of local number operators $n_1=a_1^{\dagger}a_1$ and $n_2=a_2^{\dagger}a_2$. For example, the observable $J_y$ satisfies
\be
J_y=e^{i(\pi/2)J_x}J_z e^{-i(\pi/2)J_x} .
\ee
One can then measure $J_y$ by sending the input state through a balanced beam-splitter and measure $\frac{1}{2}(n_1-n_2)$ on the resulting state. A similar reasoning allows to measure $J_x$ and indeed any $su(2)$ observable. More generally, any polynomial in $J_x$, $J_y$ and $J_z$ can be brought to a polynomial in $J_z$ via suitable $SU(2)$ unitaries \cite{csa}.  

These $su(2)$ operators commute with the total particle number operator $N = a^{\dagger}_1a_1 + a^{\dagger}_2a_2$, and as such it might be useful to partition the state space into sectors with different fixed values of $N$. The latter, or its expectation value $\mean{N}$, can be also seen as a resource that is connected with the total energy of the system, which is bounded in practical applications. Furthermore, when the total number of particles is bounded by some $n$, one can easily see that the most useful states are in the corresponding fixed-$n$ subspace.

We also remark that a subspace with fixed $n$ corresponds to the permutationally symmetric subspace of $n$ qubits, or equivalently to the space of a single particle with total spin $j=n/2$, which has dimension $d=n+1$. In such a space, states that are eigenstates of the spin $J_{\hat{\vec{\rm n}}}$ along a direction $\hat{\vec{\rm n}}=(\hat{\rm n}_x, \hat{\rm n}_y, \hat{\rm n}_z)$ with eigenvalue $n/2$ are called {\it SU(2)-coherent}. From those states, which are not orthogonal to each other, one can construct a resolution of the identity similar to that in \cref{eq:eterodynePOVM} which reads
\be\label{eq:su2coherentPOVM}
\frac{2j+1}{4\pi} \int_{S_2} {\mathrm d} \hat{\vec{\rm n}} \ \ketbra{\hat{\vec{\rm n}}} = \id ,
\ee
where $\ket{\hat{\vec{\rm n}}}$ denotes the state pointing in the direction $\hat{\vec{\rm n}}$ and the integral is performed on the two-dimensional (Bloch) sphere $S_2$ of a spin-$j$ particle with $j=n/2$.

In the usual MZI (cf. \cref{fig2}), an input state enters a balanced beam-splitter followed by arbitrary phase shift operations and then by a recombining beam-splitter operation, which corresponds to the unitary encoding:
\be
\ket{\Psi} \mapsto  e^{i(\pi/2)J_x} e^{i(\phi_1 n_1+\phi_2 n_2)} e^{-i(\pi/2)J_x} \ket{\Psi} .
\ee
It is easy to see that 
\be
\begin{gathered}
    \phi_1 n_1+\phi_2 n_2=\frac{\phi_1+\phi_2}{2}N+(\phi_1-\phi_2)J_z \\
    \implies e^{i(\phi_1 n_1+\phi_2 n_2)}=e^{i\Phi N}e^{-i\theta J_z} \\
    \text{with} \quad \Phi=\frac{\phi_1+\phi_2}{2} \quad \text{and} \quad \theta=\phi_2-\phi_1 .
\end{gathered}
\ee
For states with a definite value of the total number of particles $n$ the term $e^{i\Phi N}$ corresponds to a negligible global phase. Thus, the usual MZI corresponds to a single-parameter estimation scheme with generator given by $J_y$:
\be
\ket{\Psi_\theta} = e^{-i\theta J_y} \ket{\Psi_0} .
\ee 
Taking a Fock state $\ket{n,0}$ as the input corresponds to a $SU(2)$-coherent state pointing in the $\hat{\vec{\rm z}}$ direction and results in the so-called {\it Standard Quantum Limit} (SQL) (or {\it shot-noise scaling}):
\be
\begin{gathered}
    \probez=\ket{n,0}  \quad \rightarrow \quad 4\var{J_y}_0 = n \quad  \rightarrow \quad  \var{\hat \theta} \geq 1/ [4\var{J_y}_0] = 1/n ,
\end{gathered}
\ee
where the QFI is calculated as the variance of the generator $J_y$ since it is a single-parameter estimation (cf. \cref{observable}). Such a scaling is saturated, in the vicinity of a reference point $\theta=0$, by a measurement of $J_x$ and the usual error-propagation formula for the accuracy coming from the method of moments:
\begin{eqnarray}
\var{\hat \theta}=\frac{\var{J_x}_\theta}{(\partial_\theta \ex{J_x})^2}\bigg|_{\theta\simeq 0} = \frac{\var{J_x}_0}{\ex{J_z}_0^2} = \frac 1 n . \label{estimate}
\end{eqnarray}
On the other hand, using a so-called {\it twin-Fock state} as an input gives {\it Heisenberg scaling}
\be
\ket{\Psi_0}= \ket{n/2,n/2} \quad \rightarrow \ 4\var{J_y}_0 = n(n/2+1) ,
\ee
which is also saturated by the method of moments, by measuring the second-order observable $\{J_x,J_z\}$, where $\{\cdot, \cdot\}$ denotes the anti-commutator. 

The theoretically optimal state is given by the NOON state 
\be
\ket{\Psi_0}=e^{-i(\pi/2)J_x}  \left( \frac{\ket{n,0} + \ket{0,n}}{\sqrt{2}}  \right) \ \rightarrow \ 4\var{J_y}_0 = n^2 ,
\ee
which needs the measurement of observables such as $M=\kb{0,n}{n,0} + \kb{n,0}{0,n}$, which essentially corresponds to parity measurements, or in the multi-qubit language to $n$-qubit correlators like $\sigma_x^{\otimes n}$~\cite{TothApellaniz2014}.

\begin{figure}[h]
\centering
\includegraphics[width=0.50\columnwidth]{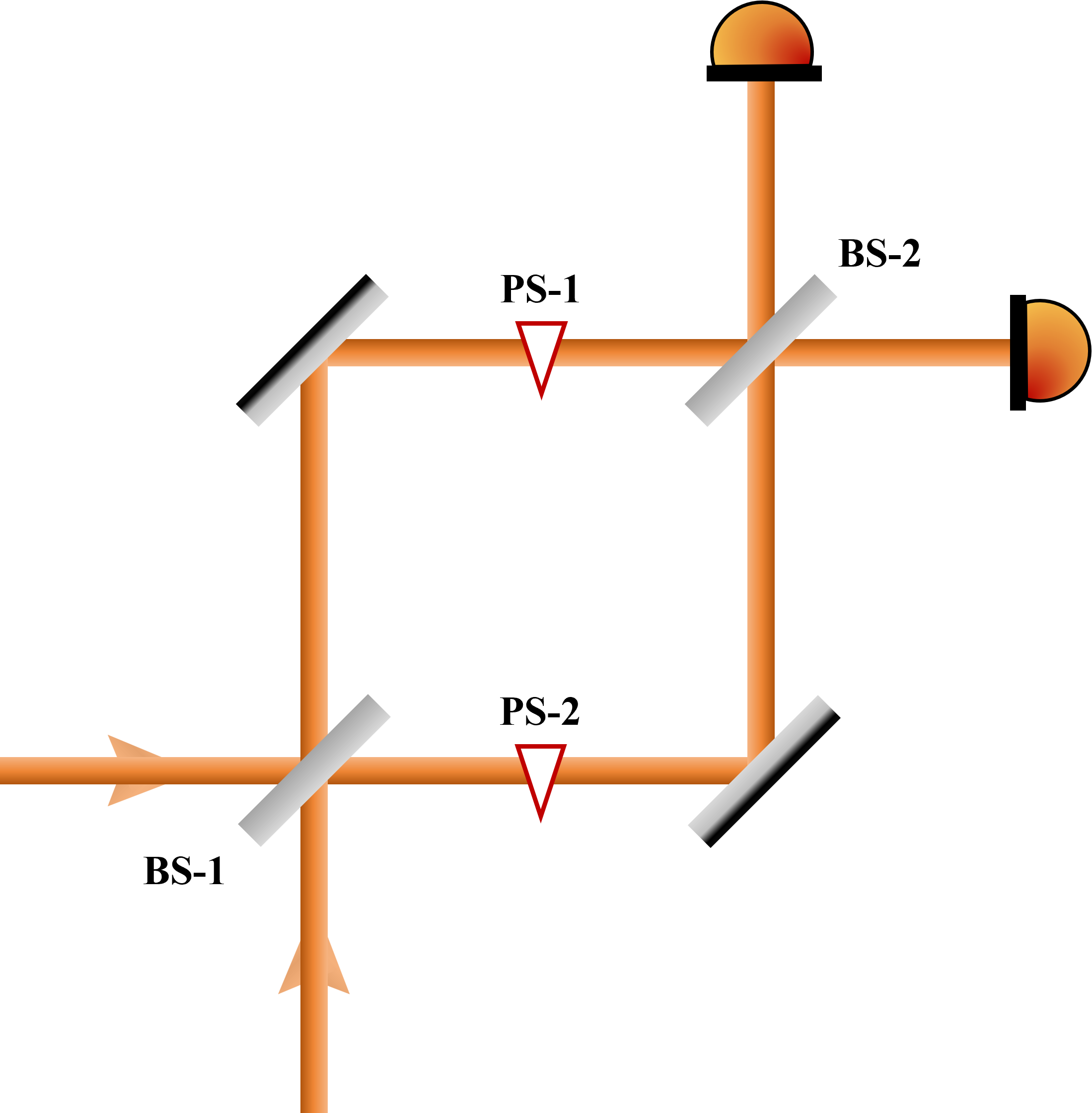} 
\caption{A Mach-Zehnder interferometer. An input state $\probez$ passes through BS-1, PS-1 and PS-2, resulting in the evolved state $\ket{\Psi_{\theta}}=U(\theta)\probez$, where $U(\theta)\in$ $SU(2)$. Measurement of spin components are then performed by sending $\ket{\Psi_{\theta}}$ through BS-2 and performing measurements of the single-mode populations $n_1$ and $n_2$.}
\label{fig2}
\end{figure}

\section{More details on three-parameter $SU(2)$ estimations with the method of moments}\label{app:detailsFullSU2}

\subsection{Local estimation and parametrizations}

Here we provide some details on the more general problem of locally estimating three $SU(2)$ phases with arbitrary values by measuring moments of the $su(2)$ generators. Beyond infinitesimal unitaries, the formulation of the problem depends on the chosen parametrization of the $SU(2)$ group. Let us first consider the fundamental representations, given by
\begin{equation}
    U(\vec \theta)=\exp(-i\frac12\sum_k \theta_k \sigma_k)=\begin{pmatrix}
        \cos(\theta/2)-i\frac{\theta_3\sin(\theta/2)}{\theta}& -\frac{\theta_2\sin(\theta/2)}{\theta}-i\frac{\theta_1\sin(\theta/2)}{\theta} \\
        \frac{\theta_2\sin(\theta/2)}{\theta}-i\frac{\theta_1\sin(\theta/2)}{\theta}&\cos(\theta/2)+i\frac{\theta_3\sin(\theta/2)}{\theta}
    \end{pmatrix} ,
\end{equation}
where $\sigma_k$ are the usual Pauli matrices and we called $\theta=\sqrt{\theta_1^2+\theta_2^2+\theta_3^2}$.

Given any element of the algebra $G\in su(2)$, and any element of the group $U \in SU(2)$, we have also the relation $U G U^{-1} \in su(2)$, which
also means that, given a basis $\{J_1,J_2,J_3\}$ of $su(2)$ we can write
\be
U G U^{-1} = \sum_k c_k J_k ,
\ee
for some coefficients $c_k$. 
In particular, a change of $su(2)$ basis is obtained as
\be
J^\prime_k = U J_k U^{-1} = \sum_l (\mathcal O_{\bf 3})_{kl} J_l ,
\ee
where $\mathcal O$ is a $3\times 3$ orthogonal matrix. 
For a generic change of basis $U(\theta)=\exp(-i\frac12\sum_k \theta_k \sigma_k)$ parametrized by three angles $\vec \theta$ such an orthogonal matrix is given by
\begin{equation}
    \mathcal O_{\bf 3}(\vec \theta)=\begin{pmatrix}
        \frac{\theta_1^2+(\theta_2^2+\theta_3^2)\cos \theta}{\theta^2}&\frac{\theta_1\theta_2(1-\cos\theta)-\theta_3\theta\sin\theta}{\theta^2} & \frac{\theta_1\theta_3(1-\cos\theta)+\theta_2\theta\sin\theta}{\theta^2}\\\frac{\theta_1\theta_2(1-\cos\theta)+\theta_3\theta\sin \theta}{\theta^2}&
        \frac{\theta_2^2+(\theta_1^2+\theta_3^2)\cos \theta}{\theta^2}
        & \frac{\theta_2\theta_3(1-\cos\theta)-\theta_1\theta\sin\theta}{\theta^2}  \\\frac{\theta_1\theta_3(1-\cos\theta)-\theta_2\theta\sin\theta}{\theta^2}&\frac{\theta_2\theta_3(1-\cos\theta)+\theta_1\theta\sin\theta}{\theta^2} & \frac{\theta_3^2+(\theta_1^2+\theta_2^2)\cos \theta}{\theta^2}
    \end{pmatrix}.
\end{equation}
A similar relation holds if we consider the generators from \cref{observable}, namely
\be
\tilde H_j(\Bth)=iU^{\dagger}(\Bth)\partial_{j} U(\Bth) = \int_0^1\mathrm{d} s\ \mathcal O_{\bf 3}(s\theta_1,s\theta_2,s\theta_3)\vec J := \tilde{\mathcal O}_{\bf 3}(\Bth) \, \vec J ,
\ee
and the matrix $\tilde{\mathcal O}_{\bf 3}(\Bth)$ can be calculated explicitly: 
\begin{equation}
    \tilde{\mathcal O}_{\bf 3}(\Bth)=\begin{pmatrix}
        \frac{\theta_1^2+(\theta_2^2+\theta_3^2)\sin\theta/\theta}{\theta^2}&\frac{\theta_1\theta_2(1-\sin\theta/\theta)-\theta_3(1-\cos\theta)}{\theta^2} & \frac{\theta_1\theta_3(1-\sin\theta/\theta)+\theta_2(1-\cos\theta)}{\theta^2}\\\frac{\theta_1\theta_2(1-\sin\theta/\theta)+\theta_3(1-\cos \theta)}{\theta^2}&
        \frac{\theta_2^2+(\theta_1^2+\theta_3^2)\sin \theta/\theta}{\theta^2}
        & \frac{\theta_2\theta_3(1-\sin\theta/\theta)-\theta_1(1-\cos\theta)}{\theta^2}  \\\frac{\theta_1\theta_3(1-\sin\theta/\theta)-\theta_2(1-\cos\theta)}{\theta^2}&\frac{\theta_2\theta_3(1-\sin\theta/\theta)+\theta_1(1-\cos\theta)}{\theta^2} & \frac{\theta_3^2+(\theta_1^2+\theta_2^2)\sin \theta/\theta}{\theta^2}
    \end{pmatrix}.
\end{equation}
Therefore, in a given parametrization $U(\vec \theta)$, the QFIM of generic $SU(2)$ unitaries of a pure state can be written as
\begin{equation}
    \mathcal F^Q[\Psi_{\vec \theta}]=4\Gamma_{\Psi_0}[\vec{\tilde H}(\Bth)]=4\tilde{\mathcal O}_{\bf 3}(\Bth)\, \Gamma_{\Psi_0}[\vec{J}] \, \tilde{\mathcal O}_{\bf 3}^T(\Bth) .
\end{equation}
In turn, this gives an asymptotic precision bound for the local estimation of $U(\vec \theta)$ around the point given by $\vec \theta$ (and again in the given parametrization).

In particular, a common parametrization that is alternative to $U(\vec \theta)=\exp[-i(\theta_x J_x+\theta_y J_y+\theta_z J_z)]$ is given by Euler angles, namely: 
\begin{equation}
    U(\Phi,\Theta,\Psi)=\exp(-i\Phi J_z)\exp(-i\Theta J_y)\exp(-i\Psi J_z),
\end{equation}
where $\vec \eta:= (\Phi,\Theta,\Psi)$ are three independent parameters just as $\vec \theta = (\theta_x,\theta_y,\theta_z)$. 
The transformation from the $\vec\theta$ representation to $\vec \eta$ reads
\begin{equation}
    \begin{cases}
    \Phi=\arctan(\frac{\frac{\theta_z}\theta\tan\frac\theta2-\frac{\theta_x}{\theta_y}}{1+\frac{\theta_x}{\theta_y}\frac{\theta_z}\theta\tan\frac\theta2})\\\Theta=2\arcsin(\frac{\sqrt{\theta_x^2+\theta_y^2}}{\theta}\sin\frac\theta2)\\\Psi=\arctan(\frac{\frac{\theta_z}\theta\tan\frac\theta2+\frac{\theta_x}{\theta_y}}{1-\frac{\theta_x}{\theta_y}\frac{\theta_z}\theta\tan\frac\theta2})
    \end{cases}.
\end{equation}
A change of parametrization then impacts the QCRB, and in particular the QFIM is transformed via the Jacobian matrix:
\begin{equation}
    \mathcal F(\Phi, \Theta,\Psi)=\mathcal J^T \, \mathcal F(\vec \theta)\, \mathcal J ,
\end{equation}
where the Jacobian matrix of the transformation $\vec \eta (\vec \theta)$ is defined as
\begin{equation}
    \mathcal J_{k\alpha}=\frac{\partial \theta_k}{\partial \lambda_\alpha},\ \ 
    k\in\{x,y,z\},\ \ \alpha\in\{\Phi,\Theta,\Psi\}.
\end{equation}
Besides considering different parametrization, one can also find a parametrization-invariant QCRB~\cite{GoldbergSanchezSotoFerrettiPRL2021}. Following Ref.~\cite{GoldbergSanchezSotoFerrettiPRL2021} one can consider the Cartan metric: 
\begin{equation}
    \vec g=\tilde {\mathcal O}_{\bf 3} \, \tilde {\mathcal O}_{\bf 3}^T=\frac{\theta^2+2\cos\theta-2}{\theta^4}\vec \theta \cdot \vec \theta^T+\frac{2(1-\cos\theta)}{\theta^2} \id_3,
\end{equation}
where $\id_3$ is the $3 \times 3$ identity matrix. Then, considering $g$ as the weight matrix on the multiparameter QCRB, one gets
\begin{equation}
    \tr[\vec g \, \COV(\Th)] \ge \frac14\tr[\Gamma_\psi^{-1}[\vec J]] ,
\end{equation} 
which is valid when the probe state is pure and is independent on the parametrization.
Applied to our investigation, we find that for the states in \cref{eq:probeXYZ}, the r.h.s is given by 
\be
\frac14\tr[\Gamma_{\Psi_k}^{-1}[\vec J]] = \frac{2}{2 k (n-k)+n}+\frac{1}{(n-2 k)^2} .
\ee

\subsection{Estimation via moments of the generators}

Let us now consider a vector of observables constructed by vectorizing the tensor
\be
A_{i_1 , \dots , i_K} = J_{i_1} J_{i_2} \dots J_{i_K} ,
\ee
where $K$ is the maximal degree of the monomial of spin observables and we consider all the different such monomials for each given order $K$. Again, working in the Heisenberg picture the moment matrix of such a vector can be written as 
\begin{equation}
    \mathcal M_{\Psi_{\vec \theta}}[\vec A]=\mathcal O(\Bth) \, \mathcal M_{\Psi_{\vec 0}}[\vec A] \,  \mathcal O^T(\Bth) ,
\end{equation} 
where $\mathcal O(\Bth)$ is an orthogonal matrix that can be constructed iteratively in the order $K$ of the monomial terms. 

For first order moments, i.e., the vector $\vec J$, we already know that the transformation is given by $\mathcal O_{\bf 3}(\Bth)$. We now illustrate how this construction works for the case $K=2$, i.e., monomials of second order. For the vector of second order monomials, which we call $\vec Q = (J_1^2,J_2^2,J_3^2,\{J_2,J_3\},\{J_1,J_3\},\{J_1,J_2\})$, we have that the orthogonal transformation matrix is given by
\begin{equation}
    \mathcal O_{\bf 6}(\Bth)=\begin{pmatrix}
        o_{11}^2&o_{12}^2&o_{13}^2&2o_{12}o_{13}&2o_{11}o_{13}&2o_{11}o_{12}\\o_{21}^2&o_{22}^2&o_{23}^2&2o_{22}o_{23}&2o_{21}o_{23}&2o_{21}o_{22}\\o_{31}^2&o_{32}^2&o_{33}^2&2o_{32}o_{33}&2o_{31}o_{33}&2o_{31}o_{32}\\o_{21}o_{31}&o_{22}o_{32}&o_{23}o_{33}&o_{22}o_{33}+o_{32}o_{23}&o_{21}o_{33}+o_{31}o_{23}&o_{21}o_{32}+o_{31}o_{22}\\o_{31}o_{11}&o_{32}o_{12}&o_{33}o_{13}&o_{12}o_{33}+o_{32}o_{13}&o_{11}o_{33}+o_{31}o_{13}&o_{11}o_{32}+o_{31}o_{12}\\o_{11}o_{21}&o_{12}o_{22}&o_{13}o_{23}&o_{12}o_{23}+o_{22}o_{13}&o_{11}o_{23}+o_{21}o_{13}&o_{11}o_{22}+o_{21}o_{12}
    \end{pmatrix},
\end{equation}
where $o_{ij}:=o_{ij}(\Bth)$ are the entries of $\mathcal O_{\mathbf{3}}$. Thus, in total, the vector $\vec A = (\vec J , \vec Q)$ transforms with the matrix $\mathcal O_{\bf 3}(\Bth) \oplus \mathcal O_{\bf 6}(\Bth)$. For higher order moments of the generators, the transformation matrix can be calculated iteratively with similar methods.

\subsubsection{Reconstructions of moments via spin-$j$ Husimi measurement}

In order to reconstruct the moments of the generators, one can employ a spin-coherent-state measurement, also known as the Husimi-$Q$ measurement, which is described by a continuous POVM on the sphere. The POVM elements are given by
\begin{equation}
E(\Omega)=\frac{2 j+1}{4 \pi}|\Omega\rangle\langle\Omega|,\qquad \Omega=(\vartheta,\varphi)\in S^2,
\qquad
\int_{S^2} E(\Omega)\,\mathrm{d}\Omega=\id ,
\end{equation}
where $\mathrm{d}\Omega=\sin\vartheta\,\mathrm{d}\vartheta\,\mathrm{d}\varphi$ and $\ket{\Omega}$ denotes the spin-$j$ coherent state pointing in the direction specified by $\Omega$. The associated unit vector is $\mathbf{n}(\Omega)=(\sin\vartheta\cos\varphi,\sin\vartheta\sin\varphi,\cos\vartheta)$. For a quantum state $\varrho_{\vec\theta}$ encoding the parameters to be estimated, the outcome distribution is the normalized Husimi function
\begin{equation}
{\rm p}(\Omega|\vec\theta)=\mathrm{tr}[\varrho_{\vec\theta}E(\Omega)]
=\frac{2j+1}{4\pi}\,\langle \Omega|\varrho_{\vec\theta}|\Omega\rangle.
\end{equation}
Choosing suitable functions $\vec f(\Omega)$ (e.g., components of the unit vector $\mathbf{n}(\Omega)$ or other spherical harmonics / symbols), one obtains $\vec\mu(\vec\theta)=\mean{\vec M(\vec \theta)}_\varrho$ for the chosen set of observables $\vec M$ directly from the Husimi statistics and can apply the method of moments as described in \cref{app:multiMoM}. In this sense, the Husimi POVM plays for spins the same conceptual role as heterodyne detection for continuous variables: it provides a single-shot, joint (non-projective) measurement whose covariances determine the asymptotic MoM precision.

To connect the continuous measurement outcomes directly to the method-of-moments employed throughout this work, we use the contravariant symbols dual to the Husimi-$Q$ POVM~\cite{GeneralizedKlimov2017}. For any operator $A$, the contravariant symbol $A^\vee(\Omega)$ is defined such that
\begin{equation}
\langle A\rangle_\varrho=\int_{S^2} \mathrm{d}\Omega\,A^\vee(\Omega)\, \tr[\varrho E(\Omega)].
\end{equation}
Given $\nu$ independent samples $\{\Omega_\ell\}_{\ell=1}^{\nu}$ from the Husimi distribution, the sample mean $\bar A_\nu=\nu^{-1}\sum_{\ell=1}^{\nu}A^\vee(\Omega_\ell)$ provides an unbiased reconstruction of $\langle A\rangle_{\varrho_{\vec\theta}}$.

For the first- and second-order moments relevant to our analysis, the contravariant symbols take explicit forms~\cite{GeneralizedKlimov2017}:
\begin{align}
(J_i)^\vee(\Omega) &= (j+1)\,n_i(\Omega), \label{eq:dual-J}\\
(\{J_i,J_k\})^\vee(\Omega) 
&= (j+1)(2j+3)\,n_i(\Omega)n_k(\Omega)-(j+1)\,\delta_{ik}. \label{eq:dual-JJ}
\end{align}
These relations allow us to map the measurement data onto the moment vector $\vec A$ used in the main text. Define the processed outcomes as $M_i(\Omega_\ell) :=(M_i)^{\vee}(\Omega_\ell)$. The sample means $\mathbb{E}_\theta\left[M_i\right]$ constructed from \cref{eq:dual-J,eq:dual-JJ} yield unbiased estimates of $\langle\vec M\rangle_{\varrho_{\vec\theta}}$, while the sample covariance matrix can be used to find the MSE.
Note that this measurement was also used in the context of estimation of three-parameter rotations in \cite{Z_Goldberg_2021}, where the classical Fisher information of this POVM was investigated. 

\end{widetext}

\bibliographystyle{quantum}

\bibliography{biblio.bib}

\end{document}